\documentclass[12pt]{article}

\usepackage{amssymb,amsfonts,amsthm}
\usepackage{amsmath}
\usepackage{bm}             
\usepackage[mathscr]{eucal}
\usepackage{cancel}
\usepackage{wasysym}

\usepackage{graphicx}
\usepackage{subfigure}
\def\be{\begin{equation}}
\def\ee{\end{equation}}
\def\bea{\begin{eqnarray}}
\def\eea{\end{eqnarray}}

\def\hsp5{\hspace{5mm}}

\theoremstyle{remark}

\newcommand{\sfrac}[2]{{\textstyle{#1\over#2}}}

\title{\sc New simple and accurate quintessence approximations}

\begin{document}

\author{
\sc Artur Alho,$^{1}$\thanks{Electronic address:{\tt
aalho@math.ist.utl.pt}}\,, and Claes Uggla,$^{2}$\thanks{Electronic address:{\tt
claes.uggla@kau.se}}\\
$^{1}${\small\em Center for Mathematical Analysis, Geometry and Dynamical Systems,}\\
{\small\em Instituto Superior T\'ecnico, Universidade de Lisboa,}\\
{\small\em Av. Rovisco Pais, 1049-001 Lisboa, Portugal.}\\
$^{2}${\small\em Department of Physics, Karlstad University,}\\
{\small\em S-65188 Karlstad, Sweden.}}


\date{}
\maketitle

\begin{abstract}

%

We derive new approximations for quintessence solutions that are simpler and
an order of magnitude more accurate than anything available in the literature, which
from an observational perspective \emph{makes numerical calculations superfluous}.
For example, our tracking quintessence approximation yields $\sim 0.1\%$ maximum
relative errors of $H(z)/H_0$ and $\Omega_\mathrm{m}(z)$ for the observationally
viable inverse power law scalar field potentials, and similarly for viable thawing
quintessence models using two slow-roll parameters. The approximations are trivially
computed from the scalar field potential and as an application we give \emph{analytic}
expressions for the CPL parameters calculated from an arbitrary scalar field potential
for thawing and tracking quintessence models.

\end{abstract}

\newpage

\section{Introduction\label{sec:intro}}

A growing number of increasingly precise cosmological observations of, e.g.,
type Ia supernovae~\cite{rieetal98,peretal99,2024dark}, the cosmic microwave
background (CMB)~\cite{planck18}, various measurements
of the Hubble parameter~\cite{2019tRGB,2021SHOES,2022Riess}, the
universe large-scale structure (LSS) and baryon acoustic oscillations
(BAO)~\cite{2020Menci,2023ApJS,2017BOSS,2020SDSS,2021eBOSS,2024desi,2024ishak,2024desivi},
suggest that the universe is presently accelerating. In addition
cosmological observations indicate that most of the matter in the
universe is invisible `dark matter'~\cite{1970Rub,1987ARA,2004Teg,keeley2024jwst}.
The standard interpretation of the observations
rely on that General Relativity is an accurate description
of gravity on cosmological scales and that the observable
Universe is almost spatially homogeneous, isotropic and flat
on sufficiently large spatial scales. This leads to that
the acceleration of the universe requires a `dark energy' (DE)
density component, $\rho_\mathrm{DE}$, with sufficiently large negative
pressure, while the dark matter seems to be best described
by a dominant cold dark matter (CDM) energy density
component, $\rho_\mathrm{m}$, without pressure. The
simplest model that is compatible
with these assumptions is the $\Lambda$CDM model, for which
the energy density $\rho_\mathrm{DE}$ and pressure
$p_\mathrm{DE}$ are given by
$\rho_\mathrm{DE} = - p_\mathrm{DE} = \Lambda >0$, where $\Lambda$
is the cosmological constant.

There are, however, theoretical problems, e.g., the nature
of dark matter~\cite{2015Wein,2017Bul}, the observed small value
of $\Lambda$ and the coincidence problem, see e.g.~\cite{peerat03} and
references therein. Recent tensions between some data
sets~\cite{AbbDES20,rieetal19,2021DiV} and that large scale structures
seem to form surprisingly early in the universe~\cite{2024Conte}
have further aggravated the situation, which has contributed
as motivation for numerous dynamical DE models for which
$\rho_\mathrm{DE}$ is evolving. Many of these models are purely
cosmographically motivated, without any tie to any theory,
notably the Chevallier-Polarski-Linder (CPL)
parametrization~\cite{chepol01,lin03}
of the DE equation of state parameter
$w_\mathrm{DE} = p_\mathrm{DE}/\rho_\mathrm{DE}$ and
a plethora of variations thereof,
see e.g.~\cite{2006Star} and references therein.

Some DE models, however, endeavour to have some theoretical motivation,
where one of the simplest attempts is quintessence, i.e. a canonical
scalar field, $\phi$, minimally coupled to gravity,
see e.g.~\cite{tsu13} and~\cite{avsetal24}
for references. This is the topic of the present paper, which
deals with a non-interacting scalar field and matter in a spatially
homogeneous, isotropic and flat spacetime, although note
that the ideas in this paper are applicable to many other models and
alternative theories. More specifically, we will \emph{derive} simple,
\emph{analytic}, and accurate approximations for thawing quintessence, which exists
for all scalar field potentials, and tracking
quintessence for asymptotically inverse power-law potentials, i.e.
$V \propto \phi^{-p}$, $p>0$ when $\phi\rightarrow 0$
(i.e. the types of quintessence that dominate the
observational literature).\footnote{Different types and nomenclatures
for quintessence are discussed in section~\ref{sec:parameters}.}
Notably, our best approximations are typically
\emph{simpler and an order of magnitude more
accurate than anything available in the literature}. For example,
our tracking quintessence approximation yields $\sim 0.1\%$ maximum relative
errors for $H(z)/H_0$ and $\Omega_\mathrm{m}(z)$ for observationally viable
inverse power law scalar field potentials, and similarly for viable thawing quintessence
models using two slow-roll parameters. We note that from an observational perspective the
remarkable accuracy of \emph{our approximations makes numerical calculations superfluous
when assessing observational constraints}; moreover the approximations yield an
analytic overview of the observational viability of broader ranges of models
than those that are typically numerically investigated.

The accuracy and scope of our approximations have far ranging consequences. There
are many papers that start from specific scalar field potentials and numerically
compute observational consequences for a variety of initial data, often not
particularly numerically systematically explored. For a wide range of scalar
field potentials the present paper makes numerical explorations mote.
Although there are mathematical considerations in this paper that might be
unfamiliar to people in this field, the results are very easy to understand.
For the reader not interested in the derivation of the approximations and
only wants to use the results, just
plug in your favourite potential into our \emph{master equations}: for \emph{thawing quintessence}
use eq.~\eqref{thawingsummary} to obtain approximations for $w_\phi(z)$, $\Omega_\phi(z)$
(or, equivalently, $\Omega_\mathrm{m}(z) = 1 - \Omega_\phi(z)$) and $H(z)$, where
$z$ is the redshift, while eq.~\eqref{w11predictions} yields
analytic expressions for the CPL parameters, thereby obtaining
simple, accurate, analytical descriptions of the recent analysis of Wolf and
Ferreira~\cite{Wolf2023} and by Shlivko and Steinhardt~\cite{shlivko2024}, where one
can also broaden the range of models; for \emph{tracking quintessence}
use eqs.~\eqref{gammabetatrac} and~\eqref{trackapprox}
to obtain approximations for $w_\phi(z)$, $\Omega_\phi(z)$
(or, equivalently, $\Omega_\mathrm{m}(z) = 1 - \Omega_\phi(z)$) and $H(z)$,
while eq.~\eqref{w11predictions} gives
analytic expressions for the CPL parameters.

The outline of the paper is as follows. The next section provides a
unifying general DE context for some of the structures that thawing
and tracking quintessence share. In Section~\ref{sec:parameters} we
derive the expressions for the parameters appearing in the previous
DE section that are particular for thawing and tracking quintessence,
respectively. In addition, we explicitly give the simple analytical
expressions for $w_\phi(z)$, $\Omega_\phi(z)$
(or, equivalently $\Omega_\mathrm{m}(z) = 1 - \Omega_\phi(z)$) and $H(z)$.
Section~\ref{sec:CPL} gives explicit analytical
expressions for the CPL parameters based on our new extremely accurate
approximations. From an observational standpoint, this eliminates the
need for numerical calculations of the CPL parameters
for whatever potential one might be interested in, which is how one
traditionally computes them for quintessence, see e.g.~\cite{Wolf2023,shlivko2024}.
Section~\ref{sec:comparison} compares the new thawing and tracking
quintessence approximations with numerical calculations and previous
approximations in the literature, reviewed in Appendix~\ref{app:quintlit}. The main
part of the paper ends with section~\ref{sec:disc}, which includes
suggestions for future applications of the present approach and results.
Appendix~\ref{app:scaling} deals with scaling freezing quintessence
(the remaining type of quintessence not covered in the main part of the paper
since $\lim_{z\rightarrow \infty}\Omega_\phi > 0$ in this case, which is
in contrast to thawing and tracking quintessence for which
$\lim_{z\rightarrow \infty}\Omega_\phi = 0$).

\section{Common thawing and tracking quintessence considerations\label{sec:DE}}

In this section we will consider rather general aspects of dark energy
that cover some features that are common for both thawing and tracking quintessence,
which we treat individually in section~\ref{sec:parameters}.

\subsection{Preliminaries\label{subsec:prel}}

Spatially homogeneous, isotropic and flat spacetimes have a line element
that can be written as\footnote{We use units such that $c=1$ and $8\pi G=1$,
where $c$ is the speed of light and $G$ is the gravitational constant.}
\begin{equation}\label{ds2N}
\begin{split}
ds^2 &= -dt^2 + a^2(t)\delta_{ij}dx^idx^j \\
     &=(1+z)^{-2}\left[H^{-2}(z)dz^2  + a_0^2\delta_{ij}dx^idx^j\right],
\end{split}
\end{equation}
where $a(t)$ is the cosmological scale factor, $a_0 = a(t_0)$
(throughout, a subscript ${}_0$ refers to quantities at the
present time $t=t_0$), where $a_0$ is usually set to one by an
appropriate scaling of the spatial coordinates;
\begin{equation}
H = \frac{\dot{a}}{a}
\end{equation}
denotes the Hubble parameter, where the dot refers to differentiation
with respect to the cosmic time $t$, while $z = (a_0/a) - 1$ is the redshift.

Since the spacetime is described by the Hubble parameter $H(z)$ it follows
that cosmological observables and their interpretation depend directly
on $H(z)$, its derivatives and integrals. This is illustrated by, e.g.,
the deceleration parameter $q = -a\ddot{a}/\dot{a}^2 = -1 + (1+z)\frac{dH}{dz}/H$,
and many distance measures, e.g., the Hubble distance $D_H(z)$,
the proper motion or coordinate distance $D_M(z)$,
the luminosity distance $D_L(z)$, the angular diameter distance $D_A(z)$,
which are given by (see, e.g.~\cite{2006Star})
\begin{subequations}
\begin{align}
D_H &= \frac{1}{H(z)},\\
D_M &= \int_0^z H^{-1}(\tilde{z})d\tilde{z},\\
D_L &= (1+z)\int_0^z H^{-1}(\tilde{z})d\tilde{z},\\
D_A &= (1+z)^{-1}\int_0^z H^{-1}(\tilde{z})d\tilde{z}.
\end{align}
\end{subequations}

To find interesting alternatives to $\Lambda$CDM some authors simply
choose some $H(z)$, e.g. $H^2$ being a polynomial in $1 + z$ as
in~\cite{2006Star}, or obtain $H(z)$ by assuming a specific form
for one of the distance measures, e.g. $D_L(z)$, which yield $H(z)$
via $d[D_L(z)/(1+z)]/dz = 1/H(z)$, see e.g.~\cite{2006Star}. More
commonly, however, is to make a choice of the DE equation of state parameter
\begin{equation}\label{wDEdef}
w_\mathrm{DE} = \frac{p_\mathrm{DE}}{\rho_\mathrm{DE}}
\end{equation}
that is different than the $\Lambda$CDM value $w_\mathrm{DE} = -1$ (see
e.g.~\cite{2021DiV} and references therein for a plethora of examples of
$w_\mathrm{DE}(z)$ and $w_\mathrm{DE}(a)$). Note, however, that $w_\mathrm{DE}$
is not a physical observable. Since cosmological observations depend on $H(z)$,
it is necessary to compute $H(z)$ from $w_\mathrm{DE}(z)$, which requires the
Einstein field equations and the matter conservation equations.

Although the distance measures are naturally expressed in terms of the
redshift $z$, computations are usually simpler using the $e$-fold time
$N = \ln(a/a_0)$, where it is subsequently trivial to transform results
to the redshift $z$ by using the relation
\begin{equation}\label{zN}
1 + z = \exp(-N).
\end{equation}

For simplicity we consider only non-interacting pressure free matter and a
dark energy component.\footnote{Although the gravitational influence of
radiation is negligible when quintessence is important for the evolution
of the Universe, we will comment on how to add it to our approximations in
section~\ref{sec:disc}.} The matter conservation equations for the matter
energy density, $\rho_\mathrm{m}>0$ and the DE energy density, $\rho_\mathrm{DE}>0$,
are given by
\begin{subequations}\label{rhoeqs}
\begin{alignat}{3}
\rho^\prime_\mathrm{m} &= -3\rho_\mathrm{m} &\quad &\Rightarrow&\quad
\rho_\mathrm{m} &= \rho_{\mathrm{m}0}\exp(-3N),\label{matterconserv} \\
\rho^\prime_\mathrm{DE} &= -3(1+w_{\mathrm{DE}})\rho_\mathrm{DE} &\quad &\Rightarrow&\quad
\rho_\mathrm{DE} &= \rho_{\mathrm{DE}0}\exp(-3N)f(N),\label{DEconserv}
\end{alignat}
\end{subequations}
where a ${}^\prime$ denotes the derivative with respect to $N$ and
where $f(N)$ is defined as
\begin{equation}\label{fdef}
f = \exp\left(-3\int_0^N w_\mathrm{DE}(\tilde{N})d\tilde{N}\right).
\end{equation}
%

\subsection{Relationships between $\Omega_\mathrm{DE}$, $\Omega_\mathrm{m}$, $w_\mathrm{DE}$ and $H$\label{subsec:par}}

Next we introduce the dimensionless and bounded Hubble-normalized variables
$\Omega_\mathrm{DE}$ and $\Omega_\mathrm{m}$:
\begin{equation}\label{dimless}
\Omega_\mathrm{DE} = \frac{\rho_\mathrm{DE}}{3H^2} = \frac{\rho_\mathrm{DE}}{\rho}, \qquad
\Omega_\mathrm{m} = \frac{\rho_\mathrm{m}}{3H^2} = \frac{\rho_\mathrm{m}}{\rho},
\end{equation}
%
%
%
where we have used the Friedmann equation (i.e. the Gauss constraint)
\begin{equation}\label{Fried}
3H^2 = \rho = \rho_\mathrm{m} + \rho_\mathrm{DE},
\end{equation}
which yields
%
\begin{equation}\label{GaussOmega}
\Omega_\mathrm{m} + \Omega_\mathrm{DE} = 1 \quad \Rightarrow \quad \Omega_{\mathrm{m}0} + \Omega_{\mathrm{DE}0} = 1.
\end{equation}
To display the relationship between these quantities and $H$ it is convenient
to introduce the present time normalized Hubble parameter
\begin{equation}
E = \frac{H}{H_0}.
\end{equation}
Equations~\eqref{Fried}, \eqref{rhoeqs}
and~\eqref{dimless} result in the following DE expression for $E^2 = E_\mathrm{DE}^2$:
\begin{equation}\label{E2}
E_\mathrm{DE}^2 = \Omega_{\mathrm{m}0}\exp(-3N) + \Omega_{\mathrm{DE}0}\exp(-3N)f(N).
\end{equation}
Using
$\rho_\mathrm{DE}/\rho_\mathrm{m} =
\Omega_\mathrm{DE}/\Omega_\mathrm{m} = \Omega_\mathrm{DE}/(1 - \Omega_\mathrm{DE})$
and~\eqref{rhoeqs}, \eqref{GaussOmega} yield
\begin{equation}\label{OmDE}
\Omega_\mathrm{DE} = \frac{\Omega_{\mathrm{DE}0}}{[\Omega_{\mathrm{m}0}/f(N)] + \Omega_{\mathrm{DE}0}}, \qquad
\Omega_\mathrm{m} = \frac{\Omega_{\mathrm{m}0}}{\Omega_{\mathrm{m}0} + \Omega_{\mathrm{DE}0}f(N)}.
\end{equation}
Since the above expressions involve $f(N)$, the integral~\eqref{fdef} needs
to be computed from $w_\mathrm{DE}(N)$.

If one instead of $w_\mathrm{DE}(N)$ has $\Omega_\mathrm{DE}(N)$
(or $\Omega_\mathrm{m}(N) = 1 - \Omega_\mathrm{DE}(N)$), then
one can obtain $w_\mathrm{DE}(N)$ and $E_\mathrm{DE}^2(N)$ from $\Omega_\mathrm{DE}(N)$ since
\begin{subequations}\label{Ombased}
\begin{align}
w_\mathrm{DE} &= \frac{1}{3}\left[\ln\left(\frac{\Omega_\mathrm{m}}{\Omega_\mathrm{DE}}\right)\right]^\prime =
\frac{1}{3}\left[\ln\left(\frac{1-\Omega_\mathrm{DE}}{\Omega_\mathrm{DE}}\right)\right]^\prime, \label{wDEOm} \\
E_\mathrm{DE}^2 &= \exp(-3N)\left(\frac{\Omega_{\mathrm{m}0}}{\Omega_\mathrm{m}}\right) =
\exp(-3N)\left(\frac{1 - \Omega_{\mathrm{DE}0}}{1 - \Omega_\mathrm{DE}}\right)\label{E2Om}
\end{align}
\end{subequations}
follows from~\eqref{rhoeqs} and~\eqref{dimless}.

We finally note that the deceleration parameter $q$ is defined as
\begin{equation}\label{qdef}
q = -\frac{a\ddot{a}}{\dot{a}^2} 
= - 1 - \frac{H^\prime}{H},
\end{equation}
and is related to the total equation of state parameter
$w_\mathrm{tot}$ according to
\begin{equation}\label{wtotq}
1 + w_\mathrm{tot} = \frac{2}{3}(1 + q),
\end{equation}
where
\begin{equation}\label{wOm}
w_\mathrm{tot} = \frac{p}{\rho} = \frac{p_\mathrm{DE}}{\rho} =
w_\mathrm{DE}\left(\frac{\rho_\mathrm{DE}}{\rho}\right) = w_\mathrm{DE}\Omega_\mathrm{DE}.
\end{equation}
To derive the above relations, we have used the Friedmann equation $\rho = 3H^2$
to obtain the last equality in~\eqref{wOm}, and also~\eqref{qdef}
and~\eqref{rhoeqs} to establish $\rho^\prime = - 3(1+w_\mathrm{tot})\rho$,
which also follows from the Einstein field equations. 

In the following sections we will discuss quintessence for which
$w_\mathrm{DE} = w_\phi$. As we will see, with the exception of
`scaling freezing quintessence' models, discussed in
Appendix~\ref{app:scaling}, the evolution of the universe
is described by `attractor' solutions for which
$\lim_{N\rightarrow - \infty}w_\phi = -1$ for thawing quintessence and
$\lim_{N\rightarrow - \infty}w_\phi = w_\infty = \mathrm{constant}$,
$-1 < w_\infty < 0$ for tracking quintessence for
asymptotically inverse power law potentials.
Thus, the evolution toward the past of these
solutions is described by the $\Lambda$CDM and $w$CDM models,
respectively. Let us therefore recall the properties of these models.


\subsection{The $\Lambda$CDM and $w$CDM models}

The $\Lambda$CDM and the $w$CDM models are characterized by $w_\mathrm{DE}=-1$
and $w_\mathrm{DE} = \mathrm{constant} = w$, respectively.
This leads to that~\eqref{fdef} results in
\begin{equation}
f = e^{3N} \quad \text{and} \quad f= e^{-3w N},
\end{equation}
respectively, where~\eqref{OmDE} and~\eqref{E2} then yield
\begin{subequations}\label{OmwEw}
\begin{xalignat}{2}
\Omega_{\Lambda} &= \frac{\Omega_{\Lambda 0}}{\Omega_{\mathrm{m}0}e^{-3N} + \Omega_{\Lambda 0}}, &\quad
\Omega_w &= \frac{\Omega_{w 0}}{\Omega_{\mathrm{m}0}e^{3w N} + \Omega_{w 0}},\label{Omw}\\
E^2_{\Lambda} &= 
\Omega_{\mathrm{m}0}e^{-3N} + \Omega_{\Lambda 0}, &\quad
E^2_{w} &= 
\Omega_{\mathrm{m}0}e^{-3N} + \Omega_{w 0}e^{-3(1+w)N},\label{LCDME1}
\end{xalignat}
\end{subequations}
where we use the subscripts $_\Lambda$ and $_{w}$ to characterize the $\Lambda$CDM and $w$CDM
expressions for $\Omega_\mathrm{DE}$ and $E_\mathrm{DE}^2$, respectively. 
%
%

\subsection{Past DE series expansions\label{subsec:pastDE}}


Equation~\eqref{wDEOm} can be written as
\begin{equation}\label{OMDEprime}
\Omega_\mathrm{DE}^\prime = -3w_\mathrm{DE}\Omega_\mathrm{DE}(1 - \Omega_\mathrm{DE}).
\end{equation}
As we will see, thawing quintessence and tracking quintessence
are associated with $\lim_{N\rightarrow - \infty}w_\phi = w_\infty = -1$
and $\lim_{N\rightarrow - \infty}w_\phi = w_\infty = \mathrm{constant} \in (-1,0)$,
respectively. To derive some common
features of these two types of quintessence, we therefore assume that
\begin{equation}\label{winfty}
w_\infty = \lim_{N\rightarrow -\infty}w_\mathrm{DE} = \mathrm{constant} \in [-1,0),
\end{equation}
which in combination with~\eqref{OMDEprime} yields
\begin{equation}\label{omdelim}
\lim_{N\rightarrow -\infty}\Omega_\mathrm{DE} = 0.
\end{equation}
Thus, $\lim_{N\rightarrow -\infty}\Omega_\mathrm{m} = 1$, i.e. the
asymptotic past is matter dominated.
Equations.~\eqref{OMDEprime}, \eqref{winfty} and~\eqref{omdelim} lead to the approximation
$\Omega_\mathrm{DE}^\prime \approx -3w_\infty\Omega_\mathrm{DE}$ when $N\rightarrow -\infty$
and hence
\begin{equation}\label{omexp}
\Omega_\mathrm{DE} \approx C e^{-3w_\infty N}.
\end{equation}

Motivated by this, let us now consider $w_\mathrm{DE}$ and $\Omega_\mathrm{DE}$ and
perform series expansions in
\begin{equation}
T = T_0e^{-3w_\infty N},
\end{equation}
where $T\rightarrow 0$ when $N\rightarrow - \infty$.
If one starts with a series expansion in $w_\mathrm{DE}$ and uses~\eqref{fdef}
and~\eqref{OmDE} one obtains an expression for $\Omega_\mathrm{DE}$
that subsequently can be series expanded in $T$; on the other hand,
if one starts with a series expansion in $\Omega_\mathrm{DE}$ and inserts
this into~\eqref{wDEOm} one obtains an expression for $w_\mathrm{DE}$
that one then can series expand in $T$. These two procedures result
in the same series expansions, which can be written as\footnote{The given
expressions can be generalized by writing
$w_\mathrm{DE} = w_\infty\left[1 + \alpha T + \sigma T^2 + \dots\right]$,
which subsequently can be used to compute the series expansions for
$\Omega_\mathrm{DE}$ and, e.g., $E_\mathrm{DE}^2$. The reason we have
written $\alpha$ as $-(\gamma - 1)$ and $\sigma$ as $(\gamma-1)\beta$ is
that this simplifies and is compatible with our quintessence approximations.
}
\begin{subequations}
\begin{align}
w_\mathrm{DE} &= w_\infty\left[1 - (\gamma - 1)T + (\gamma - 1)\beta T^2 + \dots\right] \label{wDEexp}\\
\Omega_\mathrm{DE} &= T\left(1 - \gamma T + \left[1+\frac{1}{2}(\gamma -1)\left(3 + \gamma + \beta\right)\right]T^2 + \dots\right).\label{OphiDEexp}
\end{align}
\end{subequations}
%

%
%

\subsection{DE Pad\'e approximants and present initial data\label{subsec:Pade}}

For brevity we will restrict our considerations by following the most common
approach in the literature, which is to use $w_\mathrm{DE}$ as the starting
point.
Next we will transform the series expansion for $w_\mathrm{DE}$ in $T$
to Pad\'e approximants for two reasons:
\begin{itemize}
\item[(i)] To obtain \emph{exact} DE models
that naturally generalize the $\Lambda$CDM and $w$CDM models and at the same time
form a unifying DE context for some of the structures that thawing
and tracking quintessence share. Moreover, the Pad\'e approximants
create a link between the past and the present day initial data
$\rho_{\mathrm{m} 0}$ and $H_0$ via
$\Omega_{\mathrm{DE} 0} = 1 - \Omega_{\mathrm{m} 0} = 1 - \rho_{\mathrm{m} 0}/3H_0^2$.
This section thereby forms a unifying starting point for the quintessence approximations
in the next section.
\item[(ii)] The main point of the paper: To obtain sufficiently good \emph{analytic}
approximations to the solutions of the quintessence field equations so that
numerical investigations of observational issues for thawing and tracking
quintessence become unnecessary! This is accomplished by transforming the
series expansions that approximate solutions to the quintessence equations in the
next section toward the past to Pad\'e approximants
that also are accurate at the present time.\footnote{It is quite common to use
Pad\'e approximants to improve convergence in a wide range of contexts, e.g. to obtain
accurate approximations for solutions to differential equations; for a brief review,
references, and applications in a cosmological setting, see,
e.g.,~\cite{alhugg15}.}
\end{itemize}

Our first and second order expansions in $T$ for $w_\mathrm{DE}$ yield the
following Pad\'e approximants for $w_\mathrm{DE}$, respectively, which thereby serve as
our DE models:
\begin{subequations}\label{WDET}
\begin{align}
w_\mathrm{DE} &= [0/1]_{w_\mathrm{DE}}(T) = \frac{w_\infty}{1 + (\gamma - 1)T},\label{wDE01}\\
w_\mathrm{DE} &= [1/1]_{w_\mathrm{DE}}(T) = w_\infty\left(\frac{1 + [\beta - (\gamma - 1)]T}{1 + \beta T}\right) =
w_\infty\left(1 - \frac{(\gamma - 1)T}{1 + \beta T}\right),\label{wDE11}
\end{align}
\end{subequations}
where we recall that $T = T_0e^{-3w_\infty N}$.
Note that setting $\beta=\gamma-1$ in $w_\mathrm{DE}=[1/1]_{w_\mathrm{DE}}(T)$
yields $w_\mathrm{DE} = [0/1]_{w_\mathrm{DE}}(T)$
(and similarly for the expressions below for $\Omega_\mathrm{DE}$ and $E_\mathrm{DE}$
that follow from $w_\mathrm{DE} = [1/1]_{w_\mathrm{DE}}(T)$). Furthermore, setting
$\gamma=1$ results in $w_\mathrm{DE} = w_\infty$ and hence $\Lambda$CDM when
$w_\infty = -1$ and $w$CDM when $-1<w_\infty<0$.

To connect with initial data at the present time we can set $N=0$ and solve for $T(0)=T_0$
in~\eqref{WDET} in terms of the left hand side $w_\mathrm{DE}(0) = w_{\mathrm{DE}0}$. Since results obtained
from~\eqref{wDE01} can be obtained from the corresponding ones obtained from~\eqref{wDE11}
by setting $\beta = \gamma - 1$ we now focus on the latter. It follows from~\eqref{wDE11}
and~\eqref{fdef} that the $[1/1]_{w_\mathrm{DE}}(T)$ approximant yields
\begin{subequations}\label{w11results}
\begin{align}
T_0 &= \frac{\frac{w_\infty}{w_{\mathrm{DE}0}}-1}{\beta-\frac{w_\infty}{w_{\mathrm{DE}0}}\left[\beta-(\gamma-1)\right]},\label{T0w}\\
w_\mathrm{DE} &= w_\infty \left(1-\frac{(\gamma-1) \left(\frac{w_\infty}{w_{\mathrm{DE}0}}-1\right)
	e^{-3w_\infty N}} {\beta-\frac{w_\infty}{w_{\mathrm{DE}0}}\left[\beta-(\gamma-1)\right] + \beta \left(\frac{w_\infty}{w_{\mathrm{DE}0}}-1\right)
	e^{-3w_\infty N}}\right),\\
f &= e^{-3w_\infty N}\left(\frac{\left(\gamma-1\right)\frac{w_\infty}{w_{\mathrm{DE}0}}}
{\beta-\frac{w_\infty}{w_{\mathrm{DE}0}}\left[\beta-(\gamma-1)\right] + \beta\left(\frac{w_\infty}{w_{\mathrm{DE}0}}-1\right)e^{-3w_\infty N}}
\right)^{\frac{\gamma-1}{\beta}},
\end{align}
\end{subequations}
where $f$ yields $\Omega_\mathrm{DE}$, $\Omega_\mathrm{m}$ and $E_\mathrm{DE}$ via~\eqref{OmDE} and~\eqref{E2}.
As pointed out previously, the corresponding expressions obtained from the $[0/1]_{w_\mathrm{DE}}(T)$ approximant
are conveniently obtained from the results in~\eqref{w11results} by setting $\beta = \gamma -1$.

We note, however, that $w_{\mathrm{DE}0}$ is not an observable. It is determined by $\Omega_{\mathrm{DE}0}$
and $\left.\Omega_\mathrm{DE}^\prime\right|_{N=0}$ (or $\left.\frac{\Omega_\mathrm{DE}}{dz}\right|_{z=0})$) via
eq.~\eqref{wDEOm}, which is not an easy observational task. For this reason we will instead
relate $T_0$ to $\Omega_{\mathrm{DE}0}$ and thereby obtain a direct connection to
the observables $H_0$ and $\rho_{\mathrm{m}0}$ since
$1 - \Omega_{\mathrm{DE}0} = \Omega_{\mathrm{m}0} = \rho_{\mathrm{m}0}/3H_0^2$.
We therefore consider the
$[1/1]_{\Omega_\mathrm{DE}}(T)$ Pad\'e approximant for $\Omega_\mathrm{DE}$
(use the expression~\eqref{OphiDEexp} up to second order in $T$):
\begin{equation}\label{Om11}
\Omega_\mathrm{DE} = [1/1]_{\Omega_\mathrm{DE}}(T) = \frac{T}{1 + \gamma T},
\end{equation}
from which it follows that $\Omega_{\mathrm{DE}0} = T_0/[1 + \gamma T_0]$,
which results in
\begin{equation}\label{T0}
T_0 =  \frac{\Omega_{\mathrm{DE}0}}{1 - \gamma\Omega_{\mathrm{DE}0}}.
\end{equation}
This is then inserted into~\eqref{wDE01}, which when used in~\eqref{fdef} to obtain
$f(N)$ leads to the same expression as~\eqref{Om11} with~\eqref{T0} inserted.
At the prize of additional complexity, one can continue to produce Pad\'e approximants
for $\Omega_\mathrm{DE}$ that also involve the higher order term with $\beta$ in the expansion.\footnote{The
next two higher order Pad\'e approximants, which involve $\beta$, are given by
$[1/2]_{\Omega_{\mathrm{DE}}}(T) = \frac{T}{1+\gamma T -\frac{1}{2}(\gamma-1)\left[1+(\beta-\gamma)\right]T^2}$
and $[2/1]_{\Omega_\mathrm{DE}}(T) = \frac{\gamma T+\frac{1}{2}(\gamma-1)\left[1+(\beta-\gamma)\right]T^2}
{\gamma +\left[1+\frac{1}{2}(\gamma-1)\left(3+\beta+\gamma\right)\right]T}$.}
This, however, leads to that expressing $T_0$ in terms of $\Omega_{\mathrm{DE}0}$
requires solving quadratic equations in $T_0$, which leads to rather
messy expressions. The approximation~\eqref{Om11} turns out to be a fairly good approximation
for quintessence. For this reason we will use the simple result in~\eqref{T0} for $T_0$
in~\eqref{wDE11}. This leads to that the expressions for $w_\mathrm{DE}$ in~\eqref{WDET}
can be written as follows:\footnote{Note that one can use the series expansion in $T$
for $\Omega_\mathrm{DE}$ to obtain the following expression for $E_\mathrm{DE}^2$:
$E_\mathrm{DE}^2 \approx \Omega_{\mathrm{m}0}\exp(-3N)
[1 + T - (\gamma - 1)T^2(1 + \sfrac12(\gamma - 1 + \beta)T)+ \dots]$ and use~\eqref{T0} for
$T_0$. This expression can subsequently be used to obtain various Pad\'e approximants
to obtain good quintessence approximations for $E_\phi(N)$ and then use
$E_\phi(N)$ to compute, e.g., $w_\phi(N)$, if one is so inclined.\label{for:E2}}
\begin{subequations}\label{wDEend}
\begin{align}
w_\mathrm{DE} &= [0/1]_{w_\mathrm{DE}}(T) = w_\infty\left(1 -
\frac{(\gamma-1)\Omega_{\mathrm{DE}0}}{(\gamma - 1)\Omega_{\mathrm{DE}0} + (1 - \gamma\Omega_{\mathrm{DE}0})e^{3w_\infty N}}\right),
\label{wDE01OM}\\
w_\mathrm{DE} &= [1/1]_{w_\mathrm{DE}}(T) =
w_\infty\left(1 - \frac{(\gamma-1)\Omega_{\mathrm{DE}0}}{\beta\Omega_{\mathrm{DE}0} + (1 - \gamma\Omega_{\mathrm{DE}0})e^{3w_\infty N}}\right),
\label{wDE11Om}
\end{align}
\end{subequations}
which result in the following present time values
\begin{subequations}\label{wDEPade0}
\begin{align}
w_{\mathrm{DE}0} &=  w_\infty\left(
\frac{1 - \gamma\Omega_{\mathrm{DE}0}}{\Omega_{\mathrm{m}0}}\right),\label{wDE001}\\
w_{\mathrm{DE}0} &= w_\infty\left(1 - \frac{(\gamma-1)\Omega_{\mathrm{DE}0}}{1 + (\beta - \gamma)\Omega_{\mathrm{DE}0}}\right).\label{wDE011}
\end{align}
\end{subequations}
%

Inserting the expressions for $[0/1]_{w_\mathrm{DE}}(T)$ and
$[1/1]_{w_\mathrm{DE}}(T)$ in~\eqref{WDET} and~\eqref{wDEend} into~\eqref{fdef}
yields
\begin{subequations}
\begin{align}
\begin{split}
f &= \frac{T}{T_0}\left(\frac{1 + (\gamma-1)T_0}{1 + (\gamma-1)T}\right)
= \left(\frac{\Omega_{\mathrm{m}0}}{\Omega_{\mathrm{DE}0}}\right)\left(\frac{T}{1 + (\gamma - 1)T}\right) \\
&= \frac{\Omega_{\mathrm{m}0}e^{-3w_\infty N}}
{1 - \gamma\Omega_{\mathrm{DE}0} + (\gamma-1)\Omega_{\mathrm{DE}0}e^{-3w_\infty N}}, \label{f01}
\end{split}
\\
f &= \frac{T}{T_0}\left(\frac{1 + \beta T_0}{1 + \beta T}\right)^{\frac{\gamma-1}{\beta}}
= e^{-3w_\infty N}\left(\frac{1 + (\beta-\gamma)\Omega_{\mathrm{DE}0}}
{1 - \gamma\Omega_{\mathrm{DE}0} + \beta\Omega_{\mathrm{DE}0}e^{-3w_\infty N}}\right)^{\frac{\gamma-1}{\beta}},\label{f11}
\end{align}
\end{subequations}
where the first $f$ comes from $[0/1]_{w_\mathrm{DE}}(T)$ and the second from $[1/1]_{w_\mathrm{DE}}(T)$.

It follows from~\eqref{OmDE} and~\eqref{E2} that the $[1/1]_{w_\mathrm{DE}}(T)$ Pad\'e approximant
results in
\begin{subequations}\label{OmE211}
\begin{align}
\Omega_\mathrm{DE} &= \frac{\Omega_{\mathrm{DE}0}}{\Omega_{\mathrm{m}0}e^{3w_\infty N}
\left(\frac{1 + \beta T}{1 + \beta T_0}\right)^{\frac{\gamma-1}{\beta}} + \Omega_{\mathrm{DE}0}},\label{OmDE11}\\
E_\mathrm{DE}^2 &= \Omega_{\mathrm{m}0}e^{-3N} +
\Omega_{\mathrm{DE}0}e^{-3(1 + w_\infty)N}\left(\frac{1 + \beta T_0}{1 + \beta T}\right)^{\frac{\gamma-1}{\beta}}\label{E211}
\end{align}
\end{subequations}
where we recall that $T = T_0\exp(-3w_\infty N)$,
$T_0 = \Omega_{\mathrm{DE}0}/(1 - \gamma\Omega_{\mathrm{DE}0})$,
while the result for the $[0/1]_{w_\mathrm{DE}}(T)$ Pad\'e is
obtained from the above expressions by setting $\beta=\gamma-1$, which yields\footnote{Note
that by multiplying~\eqref{OmDE01} with $\gamma$ it is easy to see that the
transformation $\gamma\Omega_\mathrm{DE} \rightarrow \Omega_w$ and
$w_\infty \rightarrow w$ yields the expression in~\eqref{Omw}, which specializes
to $\Lambda$CDM when $w_\infty = -1$. In particular this leads to
$\gamma\Omega_{\mathrm{DE}0} = \Omega_{w0}$, which illustrates the danger of
interpreting and transferring data from a $w$CDM or $\Lambda$CDM context
to a different DE context.}
\begin{subequations}\label{OME201}
\begin{align}
\Omega_\mathrm{DE} &= \frac{T}{1 + \gamma T} = \frac{\Omega_{\mathrm{DE}0}}{\gamma \Omega_{\mathrm{DE}0}
+ (1-\gamma\Omega_{\mathrm{DE}0})e^{3w_\infty N}},\label{OmDE01}\\
\begin{split}
E_\mathrm{DE}^2 &= \Omega_{\mathrm{m}0}e^{-3N}\left(1 + \frac{T}{1 + (\gamma-1)T}\right)\\
&=\Omega_{\mathrm{m}0}e^{-3N}\left(1 +
\frac{\Omega_{\mathrm{DE}0}}
{(\gamma-1)\Omega_{\mathrm{DE}0} + (1 - \gamma\Omega_{\mathrm{DE}0})e^{3w_\infty N}}\right).\label{E201}
\end{split}
\end{align}
\end{subequations}
Note the simple relationship between these expression and $\Lambda$CDM and $w$CDM, which are
obtained by setting $\gamma=1$. Furthermore, eq.~\eqref{OmDE01} implies the following
relations between the models in~\eqref{OME201} and $\Lambda$CDM and $w$CDM at the present time:
$\Omega_{\mathrm{DE}0}=\Omega_{\Lambda 0}/\gamma$ when $w_\infty=-1$
and $\Omega_{\mathrm{DE}0}=\Omega_{w 0}/\gamma$ when $w_\infty\in(-1,0)$.
As we shall see, thawing quintessence yields $\gamma>1$; on the other hand,
depending on the scalar field potential, there exist tracking quintessence models
with $\gamma >1$, $\gamma=1$ and $\gamma<1$.

Finally, note that it is trivial to express the above equations in the redshift $z$,
since $z = \exp(-N) - 1$. The above may be interpreted as \emph{exact cosmographic} DE models, but this
is \emph{not} what we are interested in. Our objective is to derive \emph{approximations}
for solutions of the field equations for \emph{thawing and tracking quintessence}.
This requires deriving $\gamma$ and $\beta$ from the quintessence field equations,
which, of course, involves the scalar field potential; this is then followed by
inserting the results for $\gamma$ and $\beta$, expressed in terms of the scalar field
potential and its derivatives, into the above DE formulas, which then are transformed
from exact DE expressions to quintessence approximations.

\section{Quintessence approximations\label{sec:parameters}}

Before deriving expressions for the parameters $\gamma$ and $\beta$ for
thawing and tracking quintessence,
we begin with some background as regards quintessence. The concepts of thawing
and freezing were defined by Caldwell and Linder (2005)~\cite{callin05} as follows:
thawing is characterized by $w_\varphi\approx - 1$ where $w_\varphi$ subsequently
grows, i.e. $w_\varphi^\prime>0$, while $w_\varphi>-1$ and $w_\varphi^\prime<0$
holds for freezing. Some quintessence models go through several stages of thawing and freezing,
see e.g.~\cite{alhetal23}, and sometimes thawing quintessence and freezing quintessence
therefore refer to a quintessence model at the present time, but this is not always
the case. Quintessence is associated with that there exist particular
quintessence solutions to the field equations that attract open sets of other quintessence solutions.
Once a solution has been attracted to an `attractor' quintessence solution
it shadows it and thereby share its properties. It should be noted that these open
sets of non-attractor quintessence solutions have a pre-quintessence stage
during the matter dominated epoch before they are attracted to an attractor
quintessence solution. During this stage of evolution the scalar field is
effectively a test field and therefore \emph{not observationally relevant}.
To obtain an unambiguous nomenclature, we use the past asymptotic properties of
the attractor solutions to denote the various types of quintessence.

For \emph{all} scalar field potentials there is a 1-parameter set of thawing quintessence
attractor solutions with $\phi = \phi_* = \mathrm{constant}$, $w_\phi = -1$,
$\Omega_\mathrm{m} = 1$, $\Omega_\varphi=0$ when $N\rightarrow-\infty$;
for asymptotically inverse power law potentials with $V \propto \phi^{-p}$, $p > 0$,
when $\phi\rightarrow 0$, there is also a single tracker quintessence attractor solution with
$-1<w_\phi = -2/(2+p) < 0$, $\Omega_\mathrm{m} = 1$, $\Omega_\varphi=0$
when $N\rightarrow-\infty$. For scalar field potentials that are sufficiently
asymptotically exponentially steep there is a third type of quintessence:
scaling freezing quintessence for which $w_\phi = 0$ when $p_\mathrm{m}=0$
and $0<\Omega_\mathrm{m} <1$ when $N\rightarrow-\infty$. As a consequence this
does not yield past continuous deformations of $\Lambda$CDM or $w$CDM and we
will therefore only touch on this topic in appendix~\ref{app:scaling}.
We thereby focus on thawing quintessence and tracking quintessence, which furthermore
dominate what is observationally investigated in the literature. For more details and a
rigorous global dynamical systems description of quintessence, see,
e.g.,~\cite{alhetal23}, \cite{alhetal24} and~\cite{alhugg23}. Next we derive
dynamical systems that are adapted to obtaining the parameters $\gamma$
and $\beta$ for thawing and tracking quintessence, respectively.

\subsection{Thawing quintessence\label{subsec:thaw}}

To obtain a dynamical system that is useful for deriving expressions for
$\gamma$ and $\beta$ for thawing quintessence we introduce the following dimensionless
bounded variable\footnote{The variable
$\Sigma_\phi$ was first introduced by Coley \emph{et al.} (1997)~\cite{coletal97}
and Copeland \emph{et al.} (1998)~\cite{copetal98} whose $x$ is $\Sigma_\phi$.
Since then, $\Sigma_\phi$ (or $\phi^\prime$) is commonly used to describe
scalar fields in cosmology, see, e.g., Urena-Lopez (2012)~\cite{ure12}, equation (2.3),
Tsujikawa (2013)~\cite{tsu13}, equation (16) and Alho and Uggla  (2015)~\cite{alhugg15b},
equation (8). The reason for using the notation $\Sigma$ for the kernel is because
$\Sigma_\phi$ plays a similar role as Hubble-normalized shear, which is typically
denoted with the kernel $\Sigma$, see e.g.~\cite{waiell97}.}
\begin{equation}\label{Sigdef}
\Sigma_\phi = \frac{\phi^\prime}{\sqrt{6}},
\end{equation}
and use $({\phi},\Sigma_{\phi},\Omega_\phi)$ as the state vector together with the $e$-fold
time $N$.\footnote{The variables $u$ and $v$ in~\cite{alhetal23} yield slower convergence.}
To obtain a dynamical system for $({\phi},\Sigma_{\phi},\Omega_\phi)$ we use~\eqref{OMDEprime}
with $\Omega_\mathrm{DE} = \Omega_\phi$ and that $w_\mathrm{DE} = w_\phi$ is given by
\begin{equation}\label{wphi}
w_\phi = -1 + \frac{2\Sigma_\phi^2}{\Omega_\phi},
\end{equation}
as follows from the definitions. In addition, the (non-linear) Klein-Gordon equation
$\ddot{\varphi} =-3H\dot{\varphi} - V_{,\varphi}$, expressed in $e$-fold time and
$\Sigma_\phi$, is needed. This leads to the dynamical system
\begin{subequations}\label{Dynsysthaw}
\begin{align}
{\phi}^\prime &= \sqrt{6}\Sigma_\phi,\\
\Sigma_\phi^\prime &= - \frac32(1 + \Omega_\phi - 2\Sigma_\phi^2)\Sigma_\phi +
\sqrt{\frac32}\,\lambda({\phi})(\Omega_\phi - \Sigma_\phi^2), \label{Sigmaeqmatter2} \\
\Omega^{\prime}_\phi &= 3(\Omega_\phi - 2\Sigma_\phi^2)(1-\Omega_\phi),\label{Omphieq2}
\end{align}
\end{subequations}
where we have introduced
\begin{equation}\label{lambdadef1}
\lambda(\phi) = - \frac{V_{,\phi}}{V}.
\end{equation}

Since thawing quintessence begins in the matter dominated regime,
$\Omega_\mathrm{m} \approx 1$, when $w_\varphi\approx - 1$ initially,
it follows from~\eqref{wphi} and from $\Omega_\mathrm{m} + \Omega_\phi = 1$
that thawing quintessence begins when
$\Sigma_\phi^2 \ll \Omega_\phi \ll \Omega_\mathrm{m}$. We then note that
whenever $\lambda(\phi)$ is bounded, which always holds for some domain(s) of $\phi$
for \emph{any} scalar field potential $V(\phi)$, there is a line of matter dominated
`Friedmann-Lema\^{i}tre' fixed points in the present dynamical systems formulation
given by
\begin{equation}
\mathrm{FL}^{\phi_*}\!:\quad (\phi,\Sigma_\phi,\Omega_\phi) = (\phi_*,0,0)
\end{equation}
and hence $\Omega_\mathrm{m} = 1$, parametrized by the constant values $\phi_*$.
To determine what happens in the vicinity of $\mathrm{FL}^{\phi_*}$ we
linearize the equations at
$(\phi,\Sigma_\phi,\Omega_\phi) = (\phi_*,0,0)$, which leads to
\begin{subequations}\label{FL_Munst}
\begin{align}
\phi &\approx \phi_* + \frac{2\lambda_*}{9}C e^{3N} -
\frac23 D e^{-\frac32 N},\label{phiseries2}\\
\sqrt{6}\Sigma_\phi &= \phi^\prime \approx \frac{2\lambda_*}{3}C e^{3N} +
D e^{-\frac32 N} \label{FL_Mbarvarphiunst} \\
\Omega_\phi &\approx C e^{3N},
\end{align}
\end{subequations}
where $\lambda_* = \lambda(\phi_*)$. It follows that thawing quintessence is
associated with the unstable manifold of $\mathrm{FL}^{\phi_*}$, obtained by
setting $D=0$. More precisely, thawing quintessence models consist of the
1-parameter set of unstable manifold orbits (i.e., solution trajectories) of
$\mathrm{FL}^{\phi_*}$ and the open set of orbits that intermediately shadows these
orbits; this open set of orbits consists of the orbits that before the thawing epoch
are pushed toward $\mathrm{FL}^{\phi_*}$ by the stable manifold orbits of
$\mathrm{FL}^{\phi_*}$ and subsequently toward the unstable manifold orbits of
$\mathrm{FL}^{\phi_*}$, which therefore approximate these orbits during their
thawing stage of evolution, for details, see~\cite{alhetal23} and~\cite{alhugg23}.

To find $\gamma$ and $\beta$ we Taylor expand $\lambda(\phi)$ around
$\phi_*$ and make a series expansion in $T = T_0 e^{3N}$ for the
variables $(\phi, \Sigma_\phi, \Omega_\phi)$. The coefficients in
the expansion are found by inserting the expansions into the
dynamical system~\eqref{Dynsysthaw} and solving for them.
Truncation of the series in $T$ gives
\begin{subequations}\label{baseapprox}
\begin{align}
\phi &\approx \phi_* + \sqrt{(\gamma-1)/3}\left[T \left(1 -\frac12\sigma T\right)\right],
\label{phiT}\\
\sqrt{6}\Sigma_\phi &= \phi^\prime \approx  \sqrt{3(\gamma-1)}\left[T(1 - \sigma T)\right],
\label{phipT}\\
\Omega_\phi &\approx T\left[1 - \gamma T + \left(1 + (\gamma-1)\left(\frac{3}{2} + \sigma\right)\right)T^2
\right] \label{OphiThawExp},
\end{align}
\end{subequations}
%
%
%
%
%
where $\gamma$ and $\sigma$ are related to the slow-roll parameters
\begin{equation}\label{epseta}
\epsilon(\phi) = \frac{1}{2}\left(\frac{V_{,\phi}}{V}\right)^2 = \frac{\lambda^2}{2},\qquad
\eta(\phi) =\frac{V_{,\phi\phi}}{V} = \lambda^2-\lambda_{,\phi},
\end{equation}
computed at $\phi = \phi_*$ on $\mathrm{FL}^{\phi_*}$, i.e.,
\begin{equation}\label{epsetastar}
\epsilon_* = \epsilon(\phi_*), \qquad \eta_* = \eta(\phi_*),
\end{equation}
according to
\begin{equation}\label{gammathaw}
\gamma = 1 + \left(\frac{2}{3}\right)^3\epsilon_*,\qquad
\sigma = \frac{4}{5}\left(1+\frac{\eta_*}{6}\right);
\end{equation}
where a comparison between~\eqref{OphiDEexp} and~\eqref{OphiThawExp} results in
\begin{equation}\label{betathaw}
\beta = \frac35 - \left(\frac{2}{3}\right)^3\epsilon_* + \frac{4}{15}\eta_*.
\end{equation}
Inserting the result for $\gamma$ in~\eqref{gammathaw} into~\eqref{wDE01OM}
and~\eqref{OME201} yields our lowest order thawing quintessence approximations
for $w_\phi$, $\Omega_\phi$ and $E_\phi$, while inserting the results for
$\gamma$ in~\eqref{gammathaw} and $\beta$ in~\eqref{betathaw} into~\eqref{wDE11Om}
and~\eqref{OmE211} gives our corresponding most accurate approximations.
Note that \emph{these formulas are valid for any scalar field potential and any} $\phi_*$,
but the approximations are only accurate for observationally viable solutions,
i.e., models for which $\epsilon(\phi_*)$ and $\eta(\phi_*)$ are typically
${\cal O}(1)$ or less. Based on section~\ref{sec:DE} where we now replace the subscript
$_\mathrm{DE}$ with $_\phi$ we obtain the following approximations for thawing quintessence:
Throughout we use eq.~\eqref{T0} to connect with initial data, i.e.
\begin{equation}\label{T0Om}
T_0 =  \frac{\Omega_{\phi 0}}{1 - \gamma\Omega_{\phi 0}},
\end{equation}
where $\Omega_{\phi 0}$ is observationally determined, which leads to that there is
no error at all for our $\Omega_\phi$ approximations at the present time,\footnote{A given
value of $\Omega_{\phi 0}$ describes a surface in the state space $(\phi,\Sigma_\phi,\Omega_\phi)$.
This surface determines where $N=0$ for the various orbits in the state space.}
although there will be small errors for both $w_{\phi 0}$ and $E_{\phi 0}$
from the $w_\phi \approx [1/1]_{\mathrm{DE}}(T)$ based approximation, which we will focus on.
On the other hand, computing $w_{\phi 0}$ \emph{numerically} for one of the
solutions from the 1-parameter set of thawing  quintessence attractor solutions
for a specific scalar field potential and using eq.~\eqref{T0w} for $T_0$ results, of course,
in no deviation at all from the numerically calculated value for $w_{\phi 0}$.

Using that $w_\infty = -1$ for thawing quintessence leads to $T = T_0\exp(3 N)$
where we recall that $T_0 =  \frac{\Omega_{\phi 0}}{1 - \gamma\Omega_{\phi 0}}$
(see e.q.~\eqref{T0Om}), while~\eqref{wDE11Om} and~\eqref{OmE211} yield
our \emph{master equations for thawing quintessence}:
\begin{subequations}\label{thawingsummary}
\begin{align}
w_\phi &\approx [1/1]_{w_\mathrm{DE}}(T) = -\left(1 - \frac{(\gamma-1)T}{1 + \beta T}\right),\label{wthaw11}\\
\Omega_\phi &\approx \frac{\Omega_{\phi 0}}{\Omega_{\mathrm{m}0}e^{-3N}
\left(\frac{1 + \beta T}{1 + \beta T_0}\right)^{\frac{\gamma-1}{\beta}} + \Omega_{\phi 0}},\label{Omthaw11}\\
E_\phi^2 &\approx \Omega_{\mathrm{m}0}e^{-3N} +
\Omega_{\phi 0}\left(\frac{1 + \beta T_0}{1 + \beta T}\right)^{\frac{\gamma-1}{\beta}},\label{E2thaw11}
\end{align}
%
where we recall that
%
\begin{align}
\gamma &= 1 + \left(\frac{2}{3}\right)^3\epsilon_*,\\
\beta &= \frac35 - \left(\frac{2}{3}\right)^3\epsilon_* + \frac{4}{15}\eta_*,
\end{align}
\end{subequations}
where the slow-roll parameters $\epsilon_*$ and $\eta_*$ are given by
$\epsilon_* = \epsilon(\phi_*) = \left.\frac{1}{2}\left(V_{,\phi}/V\right)^2\right|_{\phi=\phi_*}$
and $\eta_* = \eta(\phi_*) = \left.V_{,\phi\phi}/V\right|_{\phi=\phi_*}$; the 1-parameter set
of values $\phi_*$ are the freezing values of $\phi$ during the matter dominated regime
where $\Omega_\phi \approx 0$, where each value yields a thawing quintessence attractor
solution, where these solutions collectively describe an open set of solutions that shadow them during the
epoch when quintessence affects the spacetime structure. The simpler, but less accurate
approximations, although with similar accuracy as previous approximations in the literature,
based on $w_\phi \approx [0/1]_{w_\mathrm{DE}}(T)$, are obtained by setting $\beta=\gamma-1$ in the
above expressions. Recall also, that setting $\gamma=1$, which corresponds to that
$\epsilon_*=0$, yields the $\Lambda$CDM solution. To express our approximations in
the redshift $z$, just use the simple relation $z = \exp(-N) - 1$.
The application of the above master equations is illustrated with a specific example,
a quintessential $\alpha$-attractor inflation potential, in section~\ref{sec:comparison}.

\subsection{Tracking quintessence\label{subsec:track}}

Peebles and Ratra (1988)~\cite{peerat88,ratpee88} initiated the study
of the inverse power law potential $V \propto \phi^{-p}$, $p>0$.
They obtained approximate expressions by assuming that $H$ and $\rho$
could be approximated by their values in the matter dominated regime
and that the scale factor obeyed a power law dependence in cosmic time $t$.
Subsequently Ratra and Quillen (1992)~\cite{RatQui92} obtained the
asymptotic result $w_\phi = p_\phi/\rho_\phi = -2/(2+p)$ (their eq. (5)) for
this potential. This work was then followed up by Zlatev {\it et al.}
(1999)~\cite{zlaetal99} and Steinhardt {\it et al.} (1999)~\cite{steetal99},\footnote{See
also Podario and Ratra (2000)~\cite{PodRat00} and Peebles and Ratra (2003)~\cite{peerat03}.}
who introduced the notation `tracker field' as a form of quintessence, associated
with potentials for which $\lambda\rightarrow\infty$ when $\phi\rightarrow 0$ and
the condition $\Gamma>1$, see also~\cite{tsu13}, where
\begin{equation}\label{Gammadef}
\Gamma = \frac{V\,V_{,\phi\phi}}{V_{,\phi}^2} = 1 + (\lambda^{-1})_{,\phi}.
\end{equation}
We are here considering asymptotically inverse power law potentials, characterized by
\begin{equation} \label{lambdatr0}
\lim_{\phi\rightarrow 0}\phi\lambda = p = \mathrm{constant} > 0,\qquad \lim_{\phi\rightarrow 0}\Gamma = 1 + \frac{1}{p}.
\end{equation}

Next, we introduce a dynamical system that is suitable for deriving the parameters
$\gamma$ and $\beta$ for tracking quintessence. Tracking quintessence for
asymptotically inverse power law potentials requires that in the vicinity of $\phi=0$
the potential and $\lambda$ are monotonically decreasing, i.e., when $\lambda(\phi) >0$
and $\Gamma(\phi) > 1$, as follows from~\eqref{lambdadef1} and~\eqref{Gammadef}.
We now introduce a system of equations that are based on the following new variables:\footnote{The variables
$\tilde{\phi}$ and $v$ are slightly different than the variables $\bar{\varphi}$ and $v$
in~\cite{alhetal24}, which were adapted to a global state space setting, while the present
ones are designed for fast convergence for the tracker orbit and the explicit
appearance of $\Gamma$ in the equations and results.}
\begin{equation}\label{vartrackdef}
\tilde{\phi} = \lambda^{-2}(\phi), \qquad
u = \Sigma_\phi \sqrt{\frac{2}{\Omega_\phi}} = \frac{\phi^\prime}{\sqrt{3\Omega_\phi}},\qquad
v = \lambda(\phi)\sqrt{\frac{\Omega_\phi}{3}},
\end{equation}
from which it follows, when taken together with~\eqref{wphi},
\begin{equation}\label{uvdef}
1 + w_\phi = u^2,\qquad \Omega_\phi = 3v^2\tilde{\phi}. 
\end{equation}
These variables lead to the dynamical system
\begin{subequations}\label{uvsys}
\begin{align}
\tilde{\phi}^\prime &= 6uv\tilde{\phi}(\Gamma(\tilde{\phi}) - 1),\\
u^\prime &= \frac{3}{2}\left(2-u^2\right)(v - u),\\
v^\prime &= \frac{3}{2}\left[(1-u^2)(1-3v^2\tilde{\phi}) - 2uv(\Gamma(\tilde{\phi}) - 1)\right]v,
\end{align}
\end{subequations}
where we assume that $\Gamma(\tilde{\phi})>1$ is regular in $\tilde{\phi}$.
%
%

The dynamical system~\eqref{uvsys} admits $\tilde{\phi}=0$ (and
thereby $\phi=0$) as an invariant boundary subset. As shown in~\cite{alhetal24},
in regularizing variables, like the present ones, tracking quintessence is associated with a tracker
orbit that originates from a tracker fixed point, $\mathrm{T}$, on the
$\tilde{\phi}=0$ boundary, which in the present variables is given by
\begin{equation}
\mathrm{T}:\qquad (\tilde{\phi},u,v) = (0,u_\mathrm{T},v_\mathrm{T})
= \left(0,1,1\right)\sqrt{\frac{p}{2+p}}\,.
\end{equation}
Notably $\mathrm{T}$ is a hyperbolic saddle with the `tracker orbit' as
its unstable manifold and $\tilde{\phi}=0$ being its 2D stable
manifold. This results in that an open set of nearby orbits is pushed
toward the attractor orbit by the 2D invariant stable manifold boundary set $\tilde{\phi}=0$,
leading to that the tracker orbit acts as an `attractor orbit' that describes the evolution
of the open attracted set of tracking/shadowing orbits during their quintessence evolution,
see~\cite{alhetal24} for details.

To find the parameters $w_\infty$, $\gamma$ and $\beta$ we first note
that the tracker fixed point value for
$u = u_\mathrm{T}$ gives the past asymptotic tracker value for $w_\phi$ since
\begin{equation}
w_\infty = w_\phi|_\mathrm{T} = u_\mathrm{T}^2 - 1 = \frac{p}{2 + p} - 1 = -\frac{2}{2+p},
\end{equation}
where $w_\infty$ describes the single positive eigenvalue, given by $-3w_\infty = 6/(2+p)$.
Next we proceed and make a series expansion in
\begin{equation}
T = T_0 e^{-3w_\infty N}
\end{equation}
for the variables $(\tilde{\phi},u,v)$.
Since the governing equations~\eqref{uvsys} involve
$\Gamma(\tilde{\phi})$ we need to use a truncated Taylor series of
$\Gamma(\tilde{\phi})$ at $\tilde{\phi}=0$:
\begin{equation}\label{Gammaseries}
\Gamma(\tilde{\phi}) \approx \Gamma^{(0)} + \Gamma^{(1)}\tilde{\phi} + \frac12\Gamma^{(2)}\tilde{\phi}^2 + \dots,\qquad
\Gamma^{(n)} = \left. \frac{d^n\Gamma}{d\tilde{\phi}^n}\right|_{\tilde{\phi}=0},
\end{equation}
where $\Gamma^{(0)} = 1 + p^{-1}$.

To identify $\gamma$ and $\beta$ we solve for the coefficients for the
series expanded variables $(\tilde{\phi},u,v)$ by using the equations in~\eqref{uvsys},
which when inserted into $w_\phi = u^2 - 1$
yields $w_\phi \approx w_\infty\left[1 - (\gamma - 1)T(1 - \beta T)\right]$
%
%
%
%
%
with
\begin{subequations}\label{gammabetatrac}
\begin{align}
\gamma &= 1 - w_\infty^{-1}(1 - w_\infty^2)k,\label{gammatrack}\\
\beta &= \frac{2w_\infty^2(3w_\infty - 1) +
k\left(12w_\infty^4 - w_\infty^3 - 3w_\infty^2 + 2w_\infty-1\right) + k^{(2)}}{w_\infty(12w_\infty^2 - 3w_\infty +1)},\\
k &= \frac{w_\infty - \frac23\Gamma^{(1)}}{4w_\infty^2 - 2w_\infty + 1},\\
k^{(2)} &= \frac{w_\infty\Gamma^{(2)}}{9(w_\infty + 1)k},
\end{align}
\end{subequations}
%
%
%
where we recall that $\Gamma^{(0)} = 1 + p^{-1}$, where $p>0$ came from
$V \propto \phi^{-p}$ when $\phi\rightarrow 0$, while $\Gamma^{(1)}$ and
$\Gamma^{(2)}$ where defined in~\eqref{Gammaseries}, $w_\infty = -2/(2+p)$,
and where $\tilde{\phi}$ was defined in~\eqref{vartrackdef}, i.e.,
$\tilde{\phi} = (V_{,\phi}/V)^{-2}$. In the expression for $k^{(2)}$
we assume that $k\neq 0$, since the special class of models
with $k=0$, i.e. $\Gamma^{(1)} = \frac32w_\infty$,
needs special treatment; note, however, that this class include models with
$w_\phi=\mathrm{constant} = w_\infty$, see~\cite{alhetal24} for
details.\footnote{Sahni and Starobinsky (2000)~\cite{sahsta00},
equation (121), and Urena-Lopez and Matos (2000)~\cite{uremat00}, investigated models
with the potential $V = V_0\sinh^{-p}(\nu\phi)$ and showed that these models
admit a special solution (which turns out to be the tracker orbit) for which
$w_\phi$ is constant, i.e., they yield $w$CDM spacetimes
with $-1< w_\phi = w_\infty = w <0$.}

Using~\eqref{T0Om} for tracking quintessence in
connection with~\eqref{wDE11Om} and~\eqref{OmE211} yields
the following $w_\phi \approx [1/1]_{w_\mathrm{DE}}(T)$ based
\emph{master tracking quintessence approximations}
\begin{subequations}\label{trackapprox}
\begin{align}
w_\phi &\approx [1/1]_{w_\mathrm{DE}}(T) = -\left(1 - \frac{(\gamma-1)T}{1 + \beta T}\right),\label{wtrac11}\\
\Omega_\phi &\approx \frac{\Omega_{\phi 0}}{\Omega_{\mathrm{m}0}e^{3w_\infty N}
\left(\frac{1 + \beta T}{1 + \beta T_0}\right)^{\frac{\gamma-1}{\beta}} + \Omega_{\phi 0}},\label{Omtrack11}\\
E_\phi^2 &\approx \Omega_{\mathrm{m}0}e^{-3N} +
\Omega_{\phi 0}e^{-3(1 + w_\infty)N}\left(\frac{1 + \beta T_0}{1 + \beta T}\right)^{\frac{\gamma-1}{\beta}},\label{E2track11}
\end{align}
\end{subequations}
where we recall that $T = T_0\exp(-3w_\infty N)$,
$T_0 = \Omega_{\phi 0}/(1 - \gamma\Omega_{\phi 0})$
while $\gamma$ and $\beta$ were given in the previous eq.~\eqref{gammabetatrac}.
Setting $\beta = \gamma - 1$ in~\eqref{trackapprox} in the above
expressions yields the simpler, but less accurate,
$w_\phi \approx [0/1]_{w_\mathrm{DE}}(T)$ based tracking quintessence
approximations. Again, to express our approximations in
the redshift $z$, just use the simple relation $z = \exp(-N) - 1$.
Section~\ref{sec:comparison} provides
a specific tracking quintessence example, the inverse power law potential
$V \propto \phi^{-p}$, $p>0$.

\section{The Chevallier-Polarski-Linder parametrization\label{sec:CPL}}

The so-called CPL parameters~\cite{chepol01,lin03}
parametrize $w_\mathrm{DE}$ according to
\begin{equation}\label{CPLwDEa}
w_\mathrm{DE} = w_\mathrm{CPL} := w_0 + w_a\left(1 - \frac{a}{a_0}\right) ,
\end{equation}
which can be viewed as a truncated Taylor expansion in $a/a_0$ at the present
time (i.e. at $a/a_0 = 1$), where $a_0$ is usually set to one, which when expressed in
the redshift $z$ takes the form $w_\mathrm{CPL} = w_0 + w_a\left(\frac{z}{1+z}\right)$.
Expressing the above equation in the $e$-fold time $N$ yields
\begin{equation}\label{CPLwDE}
w_\mathrm{CPL} = w_\infty^\mathrm{CPL} - w_a e^N,
\end{equation}
where $w_\infty^\mathrm{CPL} := \lim_{N\rightarrow-\infty}w_\mathrm{CPL}$,
from which it follows that
\begin{equation}
w_\infty^\mathrm{CPL} = w_0 + w_a.
\end{equation}

Inserting~\eqref{CPLwDE} into~\eqref{fdef} yields
\begin{equation}
f = e^{-3w_\infty^\mathrm{CPL} N}\cdot e^{-3w_a(1 - e^N)}, 
\end{equation}
which results in that~\eqref{OmDE} and~\eqref{E2} yield
\begin{subequations}\label{CPLOmE2}
\begin{align}
\Omega_\mathrm{CPL} &= \frac{\Omega_{\mathrm{DE}0}}{\Omega_{\mathrm{m}0}e^{3w_\infty^\mathrm{CPL} N}\cdot e^{3w_a(1 - e^N)} + \Omega_{\mathrm{DE}0}},\\
E_\mathrm{CPL}^2 &= \Omega_{\mathrm{m}0}e^{-3N} + \Omega_{\mathrm{DE}0}e^{-3(1 + w_\infty^\mathrm{CPL})N}\cdot e^{-3w_a(1 - e^N)}.
\end{align}
\end{subequations}
We note that the expressions
in~\eqref{CPLOmE2} do not have simple single series expansion since
they mix the functions $e^N$ and $e^{3w_\infty^\mathrm{CPL} N}$, which
is in contrast to our expressions that are based on the single
function $e^{-3w_\infty N}$, as is $w$CDM and $\Lambda$CDM
when $w_\infty=-1$.
%
%

Usually the CPL parameters are \emph{cosmographically}
determined by observations, but sometimes the $w_\mathrm{CPL}(N)$ model with the
parameters $w_a$ and $w_0$, where one of these parameters can be replaced with
$w_\infty^\mathrm{CPL}$, is fitted to a \emph{numerically} calculated solution of the
quintessence field equations
(due to that most scalar field potentials do not result in explicit solutions)
with $w_\phi(N)$ at some time $N_\mathrm{fit}$ by requiring that
\begin{subequations}
\begin{align}
w_\mathrm{CPL}^\prime(N_\mathrm{fit}) &= - w_a\,e^{N_\mathrm{fit}} = w_\phi^\prime(N_\mathrm{fit}),\\
w_\mathrm{CPL}(N_\mathrm{fit}) &= w_\infty^\mathrm{CPL} - w_a\,e^{N_\mathrm{fit}} = w_\phi(N_\mathrm{fit}),
\end{align}
\end{subequations}
from which we can obtain $w_a$, $w_0$ and $w_\infty^\mathrm{CPL}$ at some fitting
time $N_\mathrm{fit}$ as follows:
\begin{subequations}
\begin{align}
w_a &= -e^{-N_\mathrm{fit}}\,w_\phi^\prime(N_\mathrm{fit}),\\
w_0 &= w_\phi(N_\mathrm{fit}) - w_\phi^\prime(N_\mathrm{fit})(1 - e^{-N_\mathrm{fit}}),\\
w_\infty^\mathrm{CPL} &= w_\phi(N_\mathrm{fit}) - w_\phi^\prime(N_\mathrm{fit}),
\end{align}
\end{subequations}
which if $N_\mathrm{fit}=0$ reduces to
\begin{subequations}
\begin{align}
w_a &= -w_\phi^\prime(0),\\
w_0 &= w_\phi(0),\\
w_\infty^\mathrm{CPL} &= w_\phi(0) - w_\phi^\prime(0).
\end{align}
\end{subequations}

In stark contrast to earlier work we will now use our quintessence approximations to
\emph{analytically compute approximate observationally indistinguishable predictions}
for the CPL parameters. Using the for thawing and tracking quintessence unifying
equations in~\eqref{wDEend} for $w_\phi(N_\mathrm{fit}) \approx w_\mathrm{DE}(N_\mathrm{fit})$
result in rather complicated expressions. For brevity we therefore use~\eqref{WDET}
where we recall that $T = T_0\exp(-3w_\infty N)$,
$T_0 = \Omega_{\phi 0}/(1 - \gamma\Omega_{\phi 0})$, where we have replaced
$\Omega_{\mathrm{DE}0}$ with $\Omega_{\phi 0}$ since we now are concerned with quintessence.
This leads to that the
expression $w_\phi \approx [1/1]_{w_\mathrm{DE}}(T)$ in~\eqref{wDE11} results in
\begin{subequations}\label{wfit}
\begin{align}
w_a &= -3w_\infty\,e^{-N_\mathrm{fit}}\,X,\\
w_0 &= w_\infty - \left[1 + 3w_\infty(1 - e^{-N_\mathrm{fit}}) + \beta T_\mathrm{fit}\right]X,\\
w_\infty^\mathrm{CPL} &= w_\infty - (1 + 3w_\infty + \beta T_\mathrm{fit})X,
\end{align}
\end{subequations}
where $X$ is defined as
\begin{equation}
X = \frac{w_\infty(\gamma-1)T_\mathrm{fit}}{(1+\beta T_\mathrm{fit})^2},
\end{equation}
where $T_\mathrm{fit} = T_0\exp(-3w_\infty N_\mathrm{fit})$.
As usual, the corresponding results for the $w_\phi \approx [0/1]_{w_\mathrm{DE}}(T)$
are easily obtained from the above ones by replacing $\beta$ with $\gamma - 1$.
This leads to simpler expressions, but not sufficiently
accurate for observational indistinguishability.

At the present time as fitting time, i.e. $N_\mathrm{fit} = 0$, the equations in~\eqref{wfit} yield
\begin{subequations}\label{w11predictions}
\begin{align}
w_a &\approx \frac{-3w_\infty^2(\gamma-1)(1 - \gamma\Omega_{\phi 0})\Omega_{\phi 0}}{(1 + (\beta - \gamma)\Omega_{\phi 0})^2},\\
w_0 &\approx w_\infty - \frac{w_\infty(\gamma-1)\Omega_{\phi 0}}{1 + (\beta - \gamma)\Omega_{\phi 0}},\\
w_\infty^\mathrm{CPL} &= w_0 + w_a,
\end{align}
%
which hence constitute our \emph{analytic quintessence predictions} for the CPL parameters
at the present time when our expressions for $\gamma$ and $\beta$ for quintessence are inserted.
It follows that
\begin{equation}\label{waw0winfty11}
\frac{w_a}{w_0-w_\infty} \approx \frac{3w_\infty(1 - \gamma\Omega_{\phi 0})}{1 + (\beta - \gamma)\Omega_{\phi 0}}
= 3w_\infty - \frac{3w_\infty\beta\,\Omega_{\phi 0}}{1 + (\beta - \gamma)\Omega_{\phi 0}},
\end{equation}
\end{subequations}
for the $w_\phi \approx [1/1]_{w_\mathrm{DE}}(T)$ case. As before, the
expressions obtained from $w_\phi \approx [0/1]_{w_\mathrm{DE}}(T)$ at $N_\mathrm{fit}=0$
are obtained by replacing $\beta$ with $\gamma-1$, where we note that
$1 + (\beta - \gamma)\Omega_{\phi 0}$ then reduces to
$1 - \Omega_{\phi 0} = \Omega_{\mathrm{m}0}$.
This leads to the following simple approximate estimate (which, unfortunately, is not at
all as good as that in~\eqref{waw0winfty11})
\begin{equation}
\frac{w_a}{w_0-w_\infty} \approx \frac{3w_\infty(1 - \gamma\Omega_{\phi 0})}{\Omega_{\mathrm{m}0}}
= 3w_\infty - 3w_\infty(\gamma-1)\left(\frac{\Omega_{\phi 0}}{\Omega_{\mathrm{m}0}}\right),
\end{equation}
which thereby is linear in $\gamma$ and equal to $3w_\infty$ for $w$CDM since $\gamma=1$ in this case.
Thawing quintessence, for which $w_\infty=-1$, yields
\begin{equation}
\frac{w_a}{w_0 + 1} = -3 + 3(\gamma-1)\left(\frac{\Omega_{\phi 0}}{\Omega_{\mathrm{m}0}}\right)
= -3 + \frac89\left(\frac{\Omega_{\phi 0}}{\Omega_{\mathrm{m}0}}\right)\epsilon_*,
\end{equation}
due to~\eqref{gammathaw}.
Thus if $\epsilon_* \approx 0$ then $w_a/(w_0+1)\approx -3$, however, viable thawing quintessence
models typically admit a range of $\epsilon_*$ up to ${\cal O}(1)$, which results in a range for
$w_a/(w_0+1) \in [-3,-1]$, cf. the discussion in~\cite{Wolf2023}, which requires the
approximation in~\eqref{waw0winfty11} with $w_\infty=-1$ in order to achieve observational
indistinguishability between a numerically computed thawing quintessence solution and our
approximated one.


\section{Numerical comparisons\label{sec:comparison}}

In this section we calculate numerical relative errors for various types of
approximations for quintessence, i.e.,
\begin{equation}
\left|\frac{\Delta F}{F}\right| = \left|\frac{F_{approx} - F_{num} }{F_{num}}\right|,
\end{equation}
where we for brevity consider the relative errors for
$w_\phi(z)$, $\Omega_\mathrm{m}(z) = 1 - \Omega_\phi(z)$ and $E_\phi(z)$
(recall that $1+z = \exp(-N)$). Apart from calculating the relative errors of
our approximations for thawing and tracking quintessence, we will also calculate
relative errors for approximations found in the literature, described in appendix~\ref{app:quintlit};
in addition, we calculate the relative errors of the CPL approximation for quintessence,
described in section~\ref{sec:CPL}, for which the CPL parameters $w_0$ and $w_a$
must be calculated \emph{numerically}. This in stark contrast to our
purely \emph{analytic} approximations which yield \emph{predictions} for
$w_\phi(z)$, $\Omega_\mathrm{m}(z) = 1 - \Omega_\phi(z)$, $E_\phi(z)$ and the
CPL parameters $w_0$ and $w_a$. Furthermore, in order to illustrate that our
most accurate approximations eliminate
the need for numerical CPL quintessence calculations, we compute
\begin{equation}\label{CPLcomperrors}
\begin{split}
\Delta(F_{[0/1]}) &= \left| \left|\frac{F_{CPL} - F_{num} }{F_{num}}\right|- \left|\frac{F_{CPL} - F_{[0/1]} }{F_{[0/1]}}\right| \right|,\\
\Delta(F_{[1/1]}) &= \left| \left|\frac{F_{CPL} - F_{num} }{F_{num}}\right|- \left|\frac{F_{CPL} - F_{[1/1]} }{F_{[1/1]}}\right| \right|,
\end{split}
\end{equation}
where $F = w_\phi(z)$, $F=\Omega_\mathrm{m}(z) = 1 - \Omega_\phi(z)$ and $F=E_\phi(z)$, while the subscripts
$_{[0/1]}$ and $_{[1/1]}$ refer to our two approximations.


%
We cannot of course test all possible scalar field potentials, but we have done so
with several potentials and obtained similar results. For brevity we will here
only present the results for two specific scalar field potentials, one illustrating
relative approximation errors for thawing quintessence and one for tracking quintessence.

\subsubsection*{Thawing quintessence}

For thawing quintessence we will consider
a currently popular model, the quintessential $\alpha$-attractor inflation model
introduced in~\cite{dimowe17} and further explored in~\cite{akretal18}, \cite{akretal20}
and~\cite{alhugg23}, referred to as Exp-model I in~\cite{akretal18}, \cite{akretal20}
and as the EC potential model in~\cite{alhugg23}, given by
\begin{equation}\label{ECPOT}
V = V_- e^{-\nu(1 + \bar{\phi})},\qquad
\bar{\phi} = \tanh\left(\frac{\phi}{\sqrt{6\alpha}}\right),
\end{equation}
which results in
\begin{equation}\label{EClambda}
\lambda = \frac{\nu}{\sqrt{6\alpha}}\left(1-\bar{\phi}^2\right),\qquad
\lambda_{,\phi} = -\left(\frac{\nu}{3\alpha}\right)\bar{\phi}\left(1 - \bar{\phi}^2\right).
\end{equation}
As a consequence~\eqref{epseta} yields
\begin{equation}\label{epsetaalphaattractor}
\epsilon_* = \frac{\nu^2}{12\alpha}\left(1 - \bar{\phi}_*^2\right)^2,\qquad
\eta_* = \frac{\nu}{6\alpha}\left(1 - \bar{\phi}_*^2\right)\left[\nu\left(1 - \bar{\phi}_*^2\right) + 2\bar{\phi}_*\right],
\end{equation}
where $\bar{\phi}_*(\phi_*)$, $\phi_* \in [-\infty,\infty]$, represents a freezing value
of $\phi$ during the matter dominated epoch for the 1-parameter
set of attractor thawing solutions with model parameters $\alpha$ and $\nu$ ($V_-$ represents
an overall scaling parameter and do not appear in the dimensionless quantities $w_\phi(z)$,
$\Omega_\mathrm{m}(z) = 1 - \Omega_\phi(z)$ and $E_\phi(z)$).

The parameters $\gamma$ and $\beta$ follow trivially from~\eqref{thawingsummary}
and are given by
\begin{equation}\label{gammabetatracker}
\gamma = 1 + \frac{2\nu^2}{81\alpha}\left(1 - \bar{\phi}_*^2\right)^2,\qquad
\beta = \frac35 + \frac{4\nu}{405\alpha}\left(1 - \bar{\phi}_*^2\right)\left[2\nu\left(1 - \bar{\phi}_*^2\right) + 9\bar{\phi}_*\right],
\end{equation}
where~\eqref{thawingsummary} also yields our $w_\phi \approx [1/1]_{w_\mathrm{DE}}(T(z))$
based approximations while setting $\beta=\gamma-1$ into~\eqref{thawingsummary} first and then
the above expression for $\gamma$ into the resulting expression yields our simpler but
less accurate $w_\phi \approx [0/1]_{w_\mathrm{DE}}(T(z))$ based approximations.

To compare with numerics we have to specify the model parameters $\alpha$ and $\nu$.
We follow~\cite{alhugg23} and set them to $(\alpha,\nu)=(\frac{7}{3},128)$.
We also need to specify which one of the 1-parameter set of thawing quintessence attractor
solutions, parametrized by $\phi_*$, we consider. These models were discussed
in~\cite{alhugg23} in the context quintessential inflation, where inflationary
considerations determined the overall energy scale $V_-$, see~\cite{akretal18} and~\cite{akretal20}.
The inclusion of radiation yielded an integral, which resulted in that only one
of the thawing quintessence attractor solutions in the 1-parameter set characterized by $\phi_*$ was compatible
with the chosen values of $V_-$, $H_0$, $\Omega_{\gamma 0}$ and $\Omega_{\mathrm{m}0}$.
This solution was found by varying the frozen scalar field value, $\phi_*$,
at the line of fixed points from which the thawing quintessence attractor solutions originated,
which resulted in $\phi_*=9.594$.
In the present case where we neglect radiation
the integral does not exist and then $H_0$ and $\Omega_{\phi 0}=0.68$ do not
impose any restrictions on $\phi_*$ at the line of fixed points
$(\phi,\Sigma_-,\Omega_\phi) = (\phi_*,0,0)$ from which the thawing quintessence attractor solutions
originate. However, due to that radiation is negligible when quintessence
affects the evolution, the present approximations for e.g. $w_\phi(z)$,
$\Omega_\phi(z)$, $E_\phi(z)$ are excellent also when
radiation is included as in~\cite{alhugg23}. For this reason, although not necessary,
we choose $\phi_*=9.594$ in order to make contact with our previous work in~\cite{alhugg23}.
Inserting $\phi_*=9.594$ and $(\alpha,\nu)=(\frac{7}{3},128)$
into~\eqref{ECPOT} and~\eqref{gammabetatracker} result in
\begin{equation}
(\gamma,\beta) = (1.095,0.789),
\end{equation}
thereby making the approximation numerically explicit.
To obtain the relative errors
we numerically integrate the dynamical system~\eqref{Dynsysthaw}
and the decoupled equation for $E_\phi = H_\phi/H_0$,
\begin{equation}
E_\phi^\prime = -\frac32\left(1 + w_\mathrm{tot}\right)E_\phi =
-\frac32\left(1 + w_\phi\Omega_\phi\right)E_\phi =
-\frac32\left(1 - \Omega_\phi + 2\Sigma_\phi^2\right)E_\phi,
\end{equation}
where we have used~\eqref{qdef}, \eqref{wtotq} and~\eqref{wphi}.
We then change from $e$-fold time $N$ to the redshift $z$ via
$z = \exp(-N) - 1$.

\begin{figure}[ht!]
	\begin{center}
		\subfigure[Relative errors for $w_{\phi}$]{\label{ECPot_wde_RelErrors}
			\includegraphics[width=0.31\textwidth]{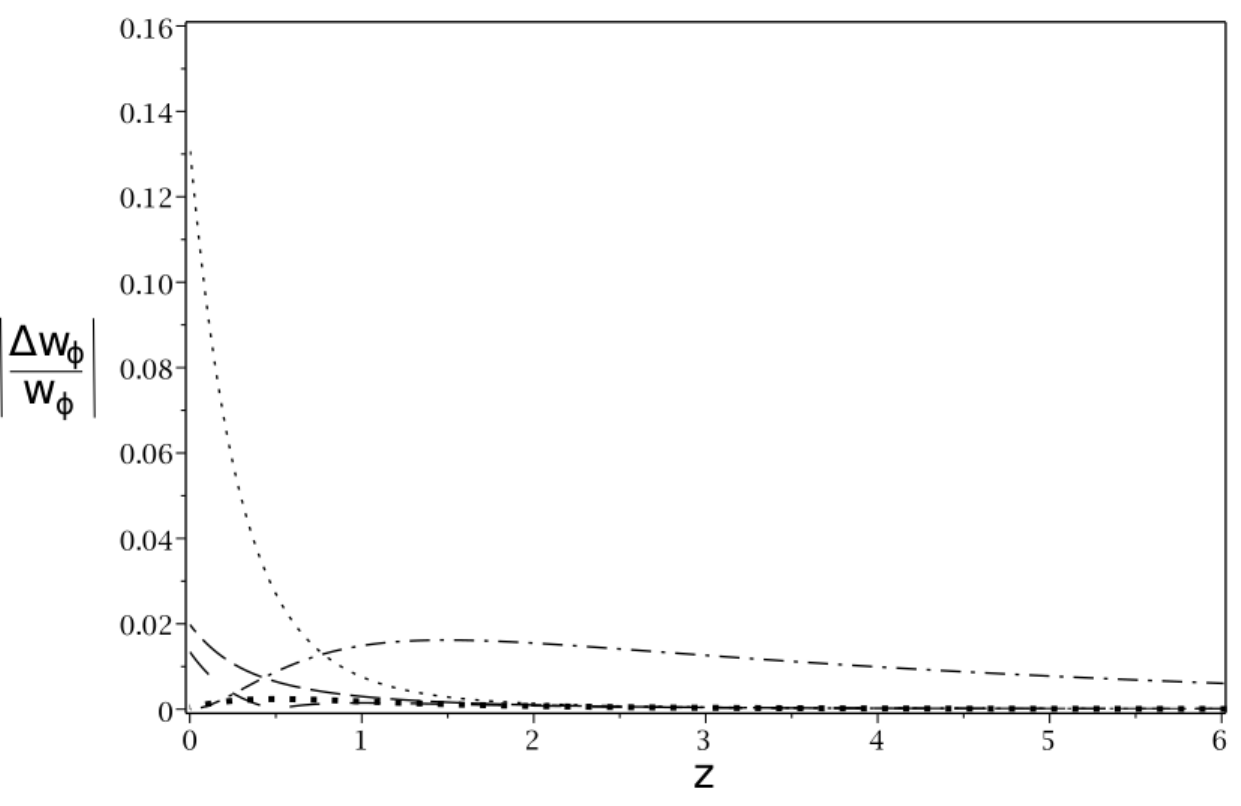}}
		\subfigure[Relative errors for $\Omega_\mathrm{m}$]{\label{ECPot_Omm_RelErrors}
			\includegraphics[width=0.31\textwidth]{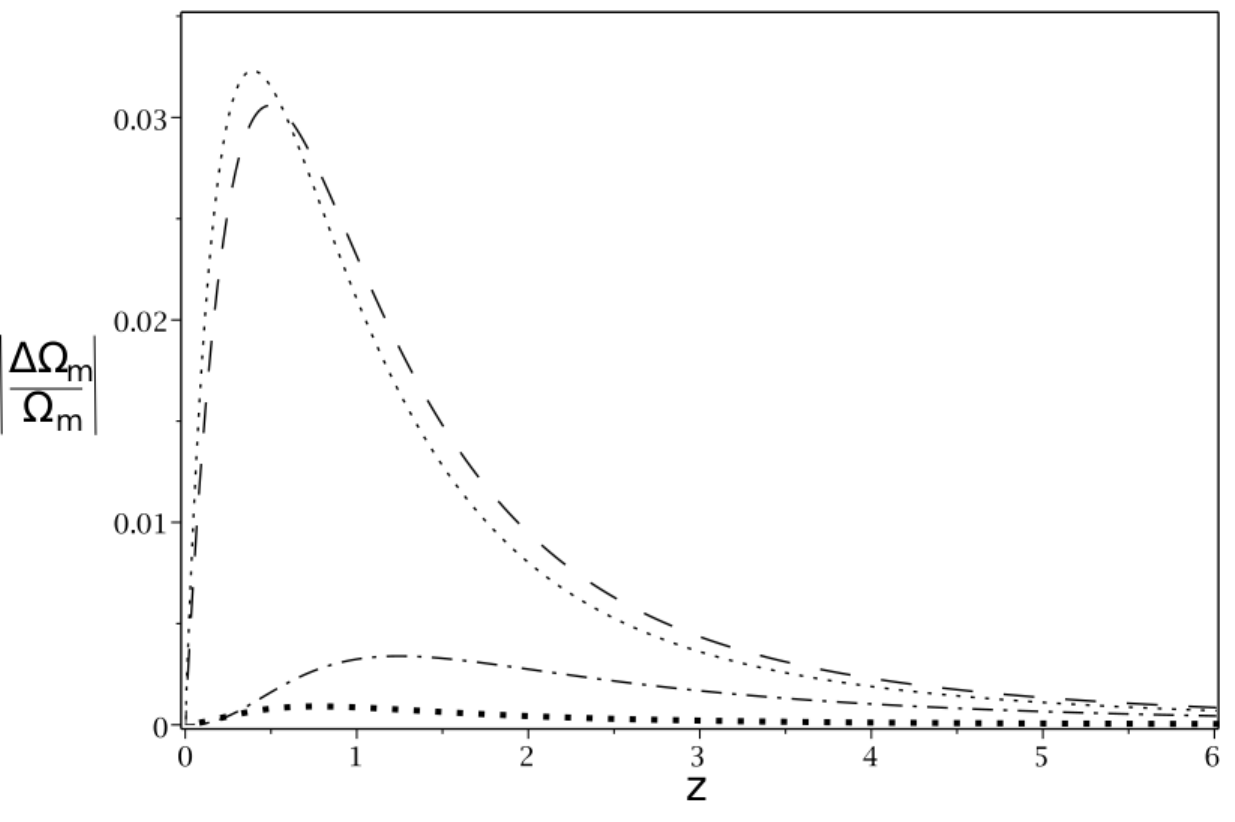}}
		\subfigure[Relative errors for $E_\phi$]{\label{ECPot_E_RelErrors}
			\includegraphics[width=0.31\textwidth]{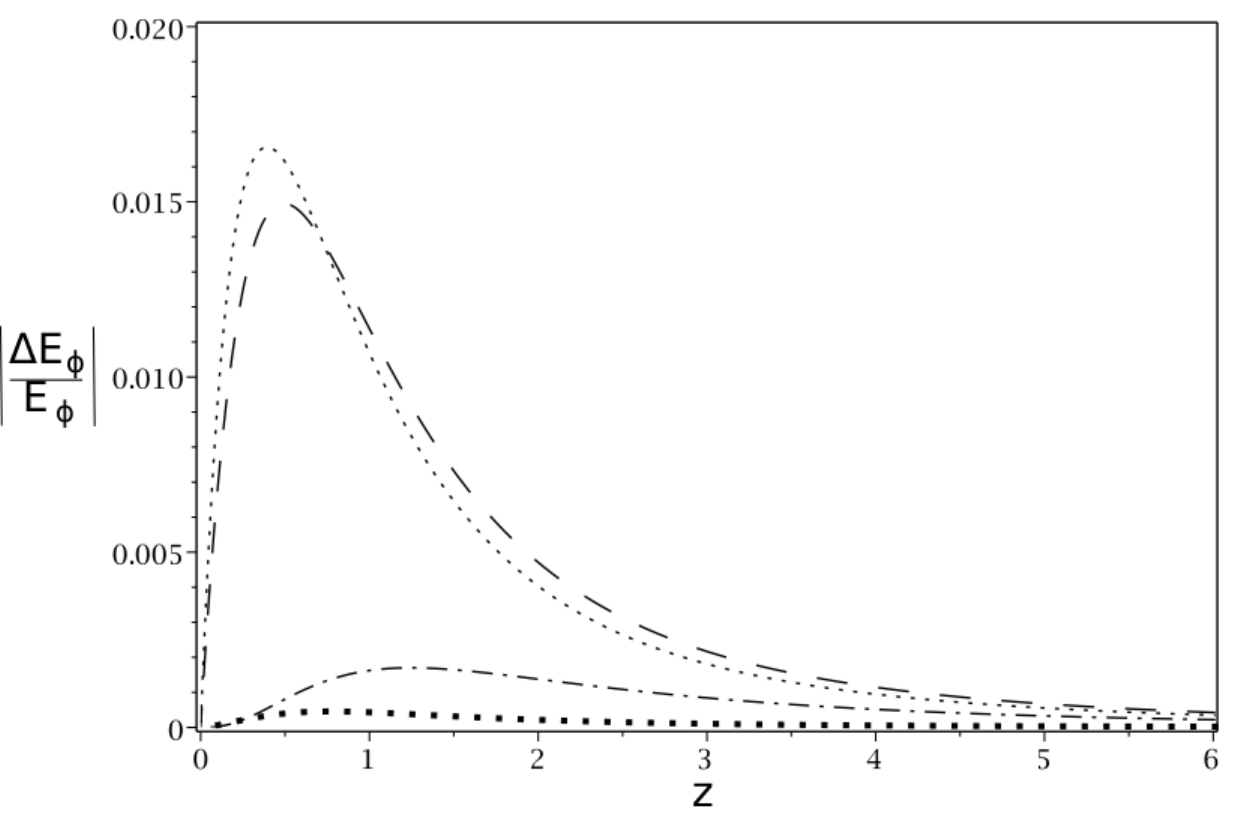}}
		\vspace{-0.5cm}
	\end{center}
	\caption{\emph{Thawing quintessence}: Relative errors $\left|\Delta F/F\right|$ for
$w_\phi(z)$, $\Omega_\mathrm{m}(z)$, $E_\phi(z)$. 
The dotted curves depict the relative errors of the new $w_\phi \approx [0/1]_{w_\mathrm{DE}}(T(z))$ based
approximations, i.e.,~\eqref{thawingsummary} with $\beta = \gamma - 1$,
while the thick dotted curves represent the relative errors of the $w_\phi \approx [1/1]_{w_\mathrm{DE}}(T(z))$
based approximations~\eqref{thawingsummary}. 
The dash-dotted curves correspond to the CPL based approximations
with the numerically determined values $(w_0,w_a)=(-0.9186,-0.0871)$.
In Figure~(a) the dashed curve is the approximation by Scherrer and Sen~\cite{schsen08}
given in eq.~\eqref{swappr}, while the space-dashed curve is the approximation obtained
by Chiba~\cite{chi09}, see eq.~\eqref{K2neg}. In Fig. (b) and (c) the dashed curves correspond to
the $\Lambda$CDM solution given in~\eqref{OmwEw}, 
which is the basis for the approximations of both Scherrer and Sen~\cite{schsen08}
and Chiba~\cite{chi09}.
}
\label{Fig_ECPotApprox3}
\end{figure}

The relative errors of $w_\phi(z)$, $\Omega_\mathrm{m}(z) = 1 - \Omega_\phi(z)$
and $E\phi(z)$ for several approximations are depicted in Figure~\ref{Fig_ECPotApprox3}.
These include analytic approximations such as our $w_\phi \approx [0/1]_{w_\mathrm{DE}}(T(z))$
and $w_\phi \approx [1/1]_{w_\mathrm{DE}}(T(z))$ based approximations, as well as
the approximations by Scherrer and Sen~\cite{schsen08} and Chiba~\cite{chi09},
given by the equations~\eqref{swappr} and~\eqref{K2neg}, respectively,
in appendix~\ref{app:quintlit}. In addition we give the CPL approximation based on
the numerically calculated values for $w_0$ and $w_a$ at $z=0$, which thereby, of course,
results in no error for $w_\phi$ at $z=0$.

As can be seen, the earlier analytic approximations by Scherrer and Sen~\cite{schsen08}
and Chiba~\cite{chi09}\footnote{Here we have treated these approximations in the same way as our
approximations, i.e. as analytic predictive approximations. This is not how they were dealt with
in the original references. There numerical calculations picked out a particular
solution in the 1-parameter set of thawing solutions, which results in less errors at the price
of loosing predictive power.}
and the more simple and consistent
$w_\phi \approx [0/1]_{w_\mathrm{DE}}(T(z))$ approximant based
approximations have similar relative errors for the physical observables $\Omega_\mathrm{m}$
and $E_\phi$, with respective $\sim 3\%$ and $\sim 1.5\%$ maxima, while the
$w_\phi \approx [1/1]_{w_\mathrm{DE}}(T(z))$ approximant yields a maximum relative error of
$\sim 0.1\%$ for $\Omega_\mathrm{m}$ and an even smaller error for $E_\phi$, making
this approximation observationally indistinguishable from numerical results. Not even the
numerically based CPL approximation, with no predictive power at $z=0$,
is as good as our analytic $w_\phi \approx [1/1]_{w_\mathrm{DE}}(T(z))$ based
approximations.

Note that if we instead of using $T_0 = \Omega_{\phi 0}/(1 - \gamma\Omega_{\phi 0})$
use the numerically computed $w_0 = w_{\phi 0}$ in $T_0$ as described in
eq.~\eqref{w11results} we get even more accurate results than our previous
approximations and the CPL based approximations, but then this results in
that the \emph{purely analytic predictions} of our quintessence approximations
are lost.

Finally, note that changing $\phi_*$ so that $\lambda_*=\lambda(\phi_*)$ decreases also decreases
the errors, which is due to slower rolling of the scalar field. Note also
that the errors of all approximations decrease as $z$ increases,
which should not come as a surprise since increasing
$z$ entails increased matter domination. Investigations of several other
scalar field potentials yield similar results.

\begin{figure}[ht!]
	\begin{center}
		\subfigure[Difference between relative CPL errors and the relative
         difference between CPL and our approximations for $w_{\phi}$]{\label{ECPot_Difwde_RelErrors}
			\includegraphics[width=0.28\textwidth]{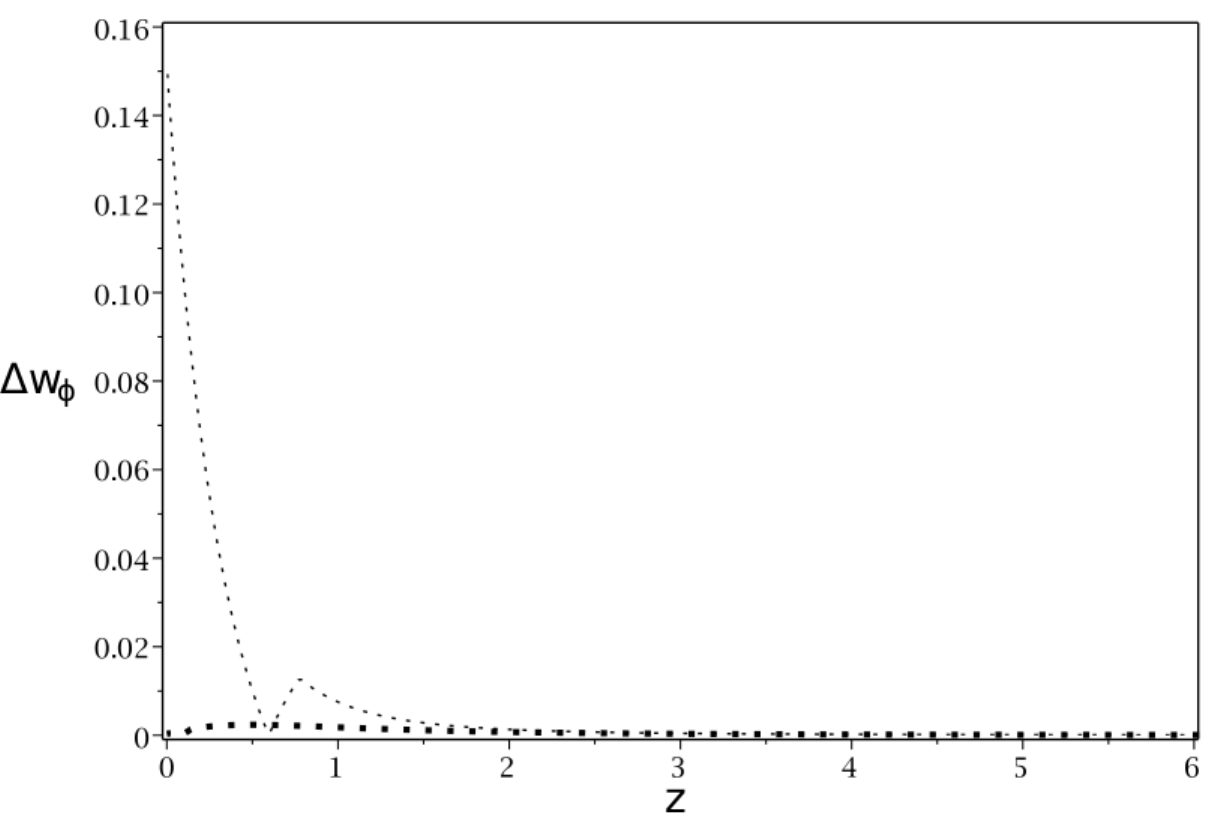}}
		\hspace{0.1cm}
		\subfigure[Difference between relative CPL errors and the relative
         difference between CPL and our approximationss for $\Omega_\mathrm{m}$]{\label{ECPot_DifOmm_RelErrors}
			\includegraphics[width=0.28\textwidth]{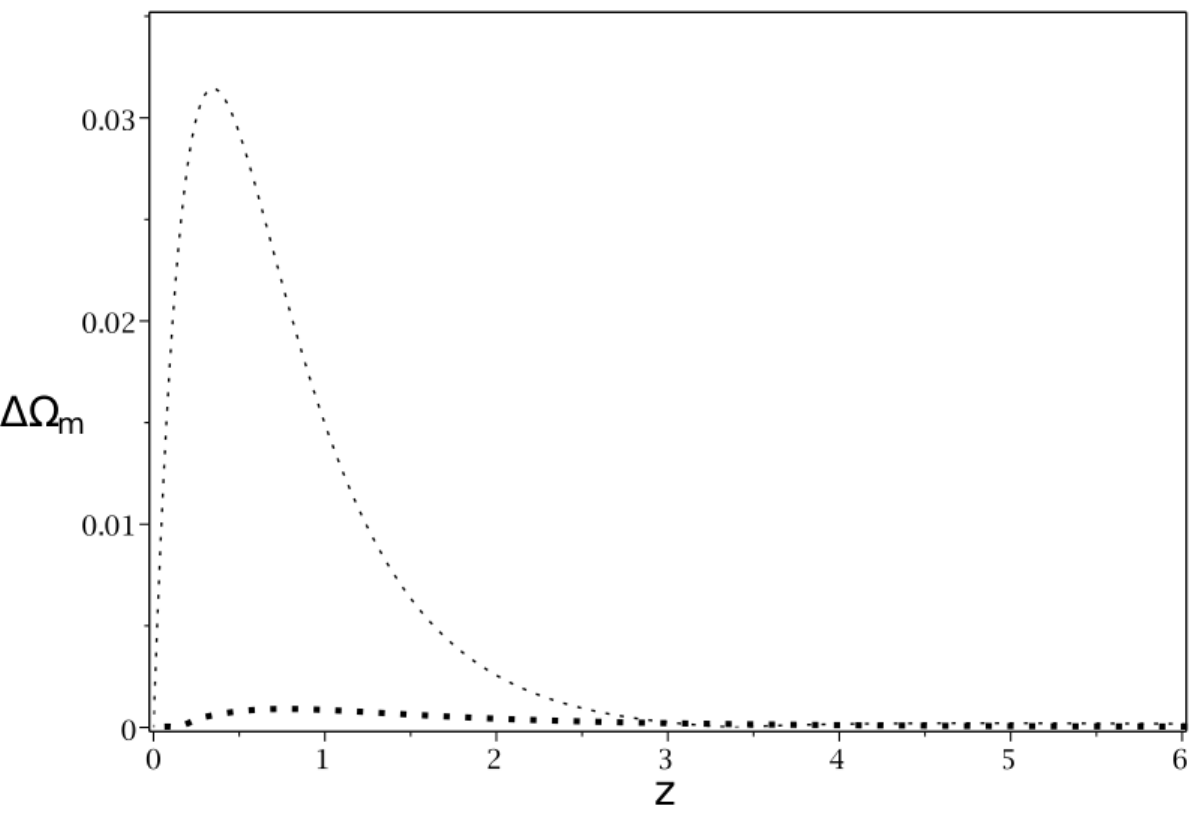}}
		\hspace{0.1cm}
		\subfigure[Difference between relative CPL errors and the relative
         difference between CPL and our approximations for $E_\phi$]{\label{ECPot_DifE_RelErrors}
			\includegraphics[width=0.28\textwidth]{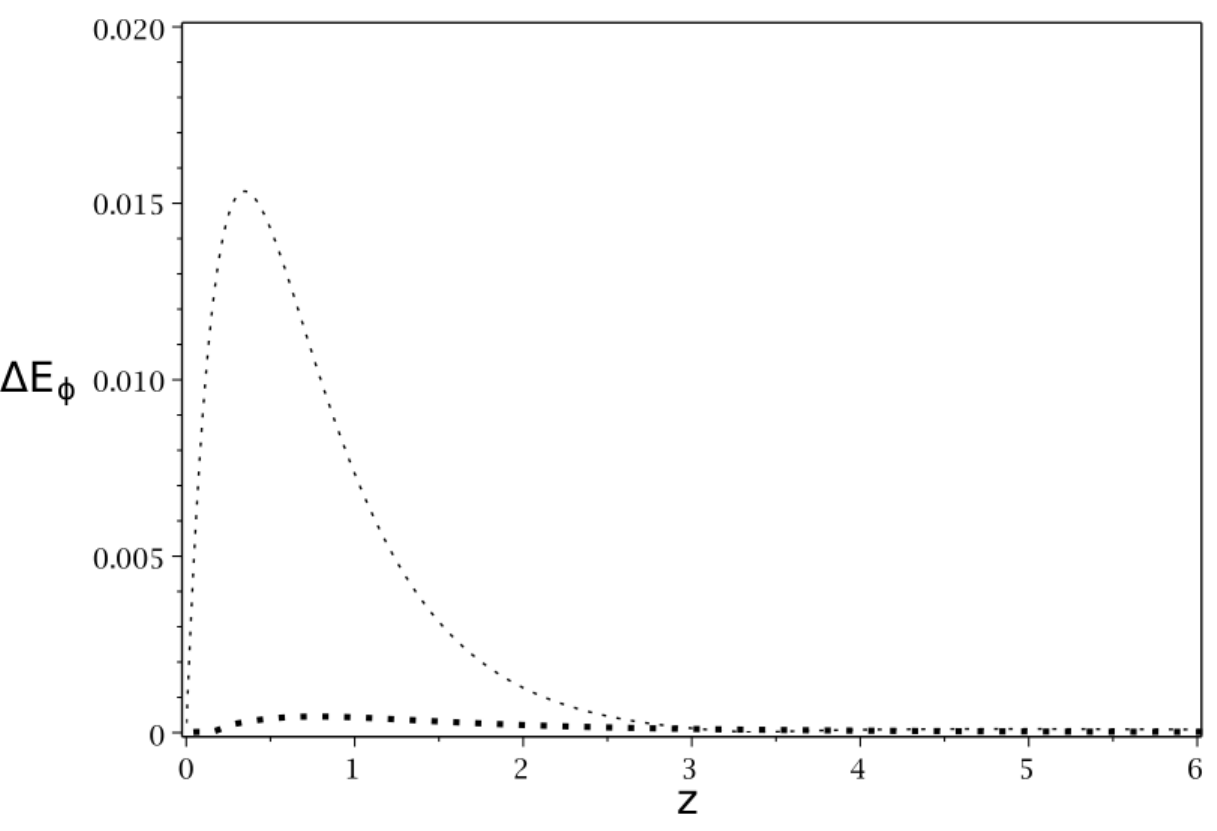}}
		\vspace{-0.5cm}
	\end{center}
	\caption{\emph{Thawing quintessence}: Difference between relative CPL errors and the relative
         difference between CPL and our approximations based on
         $w_\phi \approx [0/1]_{w_\mathrm{DE}}(T(z))$ (dotted curves) and
$w_\phi \approx [1/1]_{w_\mathrm{DE}}(T(z))$ (thick dotted curves), as described in
equations~\eqref{CPLcomperrors} for $\Delta(w_\phi(z))$, $\Delta(\Omega_\mathrm{m}(z))$, $\Delta(E_\phi(z))$,
emphasizing that the analytic $w_\phi \approx [1/1]_{w_\mathrm{DE}}(T(z))$ based approximations
are observationally indistinguishable from numerically based CPL approximations,
making such numerical explorations unnecessary.
	}
	\label{Fig_ECPotApprox}
\end{figure}

Figure~\ref{Fig_ECPotApprox} contains the differences $\Delta(w_\phi(z))$,
$\Delta(\Omega_\mathrm{m}(z))$ and $\Delta(E_\phi(z))$ between relative CPL errors
and the relative difference between CPL and our approximations, as
described in eq.~\eqref{CPLcomperrors}. The simple
$w_\phi \approx [0/1]_{w_\mathrm{DE}}(T(z))$ approximant based approximations yields order of estimate
descriptions of the CPL approximations but is not sufficiently good to make numerical calculations unnecessary.
On the other hand, our analytic $w_\phi \approx [1/1]_{w_\mathrm{DE}}(T(z))$ approximant based
approximations are virtually indistinguishable from numerical results, again illustrating that this approximation
is so good so that it makes numerical calculations irrelevant.

We finally note that our numerical calculations yield $(w_0,w_a)\approx(-0.9186,-0.0871)$, and
hence $w^{\mathrm{CPL}}_\infty \approx -1.0057$ and
$\frac{w_a}{w_0 + 1} \approx -1.0700$. Using our $w_\phi \approx [1/1]_{w_\mathrm{DE}}(T(z))$ approximant,
which yields equation~\eqref{w11predictions}, where we recall that $w_\infty = -1$ for thawing quintessence,
results in $(w_0,w_a)\approx(-0.9183,-0.0791)$ and hence $w_\infty^\mathrm{CPL} \approx -0.9974$
and  $\frac{w_a}{w_0 + 1} \approx -0.9682$, which illustrates how good this approximation is.

\subsubsection*{Tracking quintessence}

As our example for tracking quintessence we choose the inverse power law potential,
for which
\begin{equation}
V = V_0\phi^{-p}, \qquad \lambda = \frac{p}{\phi}, \qquad \Gamma = 1 + \frac{1}{p} = \mathrm{constant}.
\end{equation}
Thus the derivatives of $\Gamma$ are zero and hence
$\Gamma^{(1)}=0$ and $\Gamma^{(2)}=0$ in~\eqref{gammabetatrac}, while, as
for all the present tracking quintessence models,
\begin{equation}
w_\infty = -\frac{2}{2+p},
\end{equation}
where we will use $w_\infty$ rather than $p = -2(1 + w_\infty)/w_\infty$
since this leads to somewhat more succinct expressions.
The expressions for $\gamma$ and $\beta$ for this case follows
straightforwardly from~\eqref{gammabetatrac} and can be written as
\begin{subequations}\label{pgammabeta}
\begin{align}
\gamma &= 1 - \frac{1 - w_\infty^2}{4w_\infty^2 - 2w_\infty + 1},\\
\beta &= \frac{36w_\infty^4 - 21w_\infty^3 + 7w_\infty^2 - 1}{(4w_\infty^2 - 2w_\infty + 1)(12w_\infty^2 - 3w_\infty + 1)}.
\end{align}
\end{subequations}
which for $p=1$ yields $w_\infty = -2/3$ and $(\gamma,\beta) \approx (0.86486,0.45081)$,
while, e.g., $p=0.1$ results in $w_\infty = -20/21 \approx -0.95238$ and
$(\gamma,\beta) \approx (0.98577,0.55145)$.
Inserting these results into~\eqref{trackapprox} yields our tracking quintessence approximations
(as before, setting first $\beta=\gamma-1$
yields our $w_\phi \approx [0/1]_{w_\mathrm{DE}}(T(z))$ based approximations).
To make numerical comparisons with our approximations we choose the
historically important case $p=1$ described above, though this value
might be too large to constitute an observationally viable model.

\begin{figure}[ht!]
	\begin{center}
		\subfigure[Relative errors of $w_{\phi}$]{\label{RPPot_wde_RelErrors}
			\includegraphics[width=0.3\textwidth]{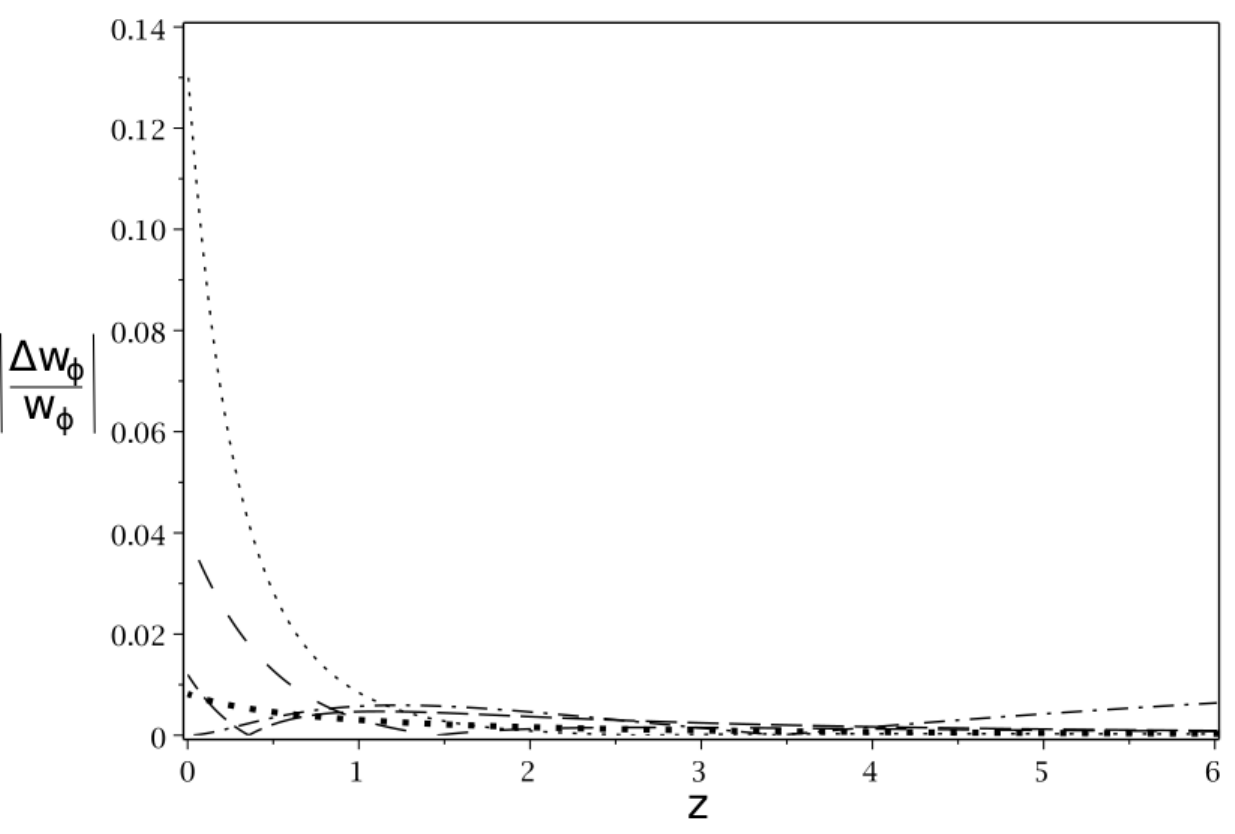}}
		\hspace{0.1cm}
		\subfigure[Relative errors of $\Omega_\mathrm{m}$]{\label{RPPot_RelError}
             \includegraphics[width=0.3\textwidth]{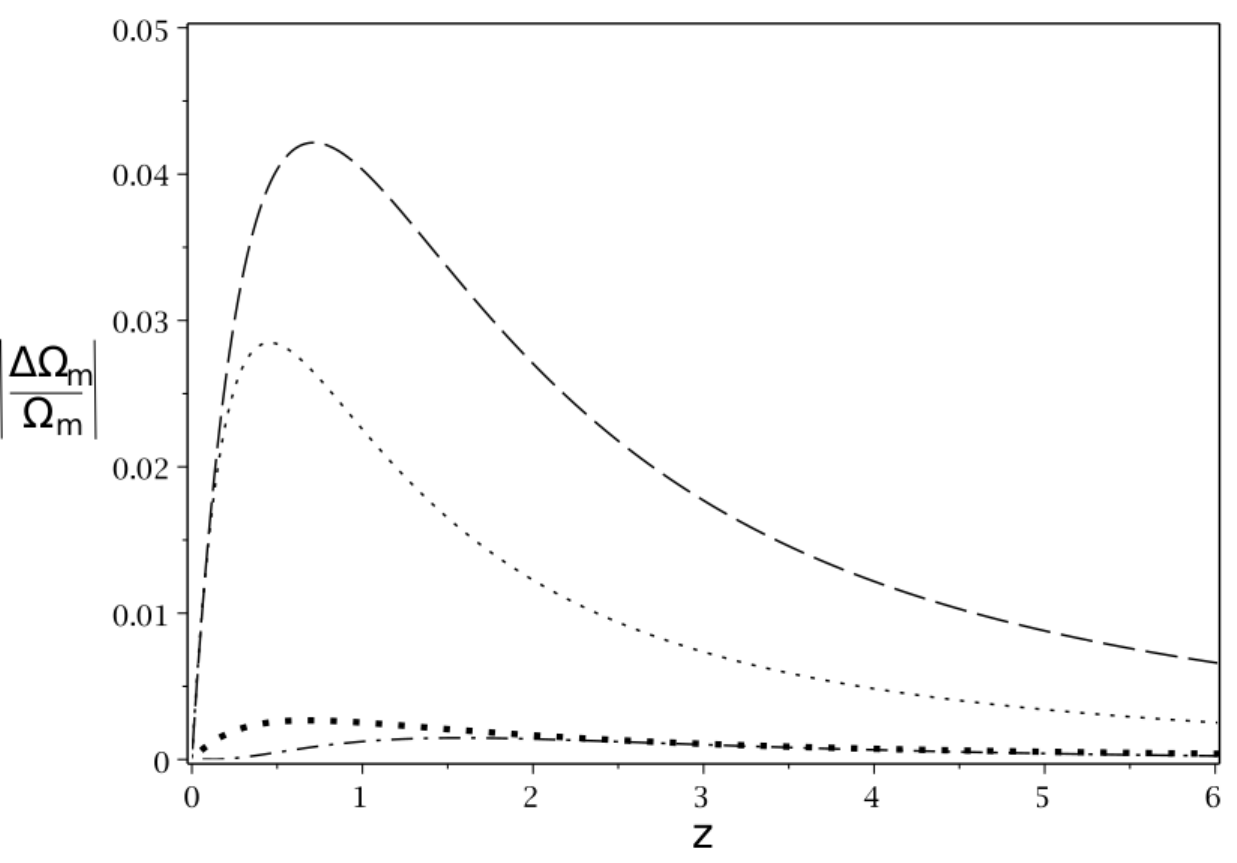}}
		\hspace{0.1cm}
		\subfigure[Relative errors of $E_\phi$]{\label{RPPot_E_RelErrors}
			\includegraphics[width=0.3\textwidth]{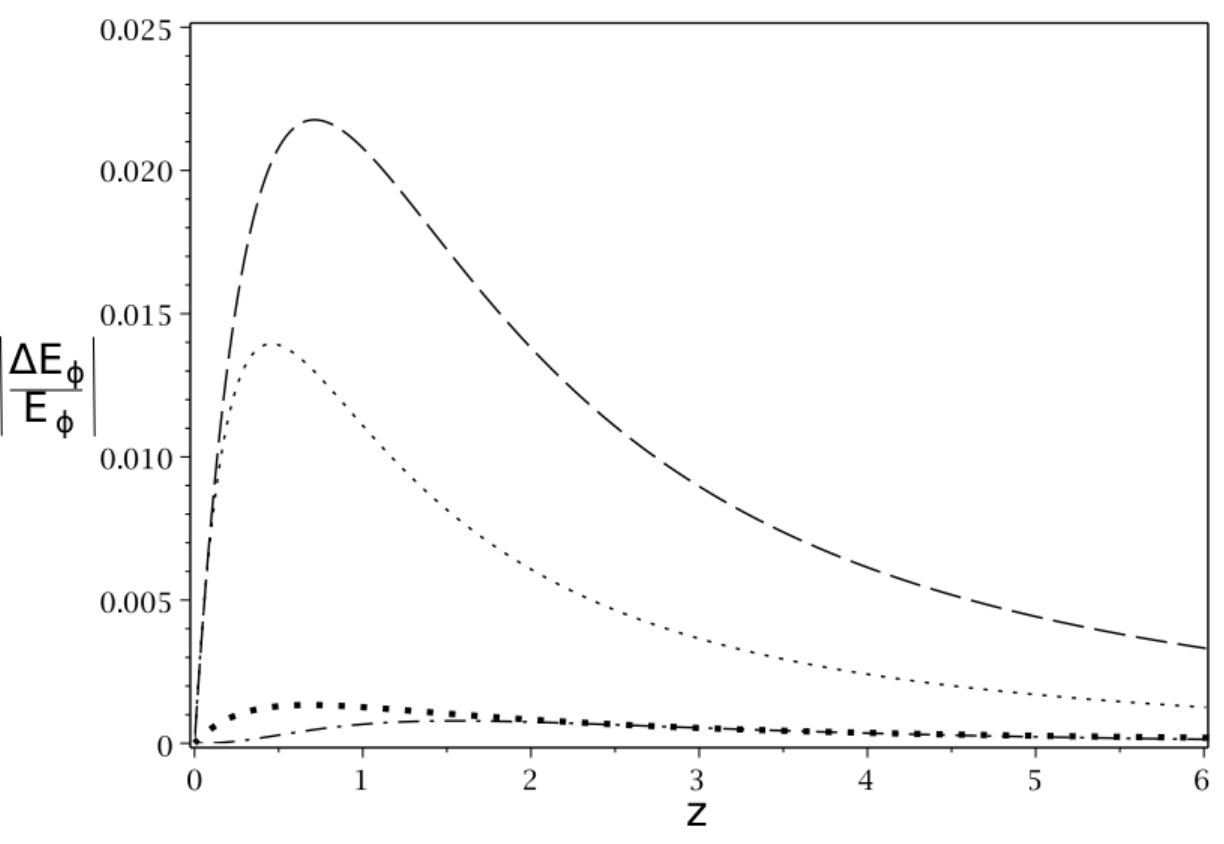}}
		\vspace{-0.5cm}
	\end{center}
	\caption{\emph{Tracking quintessence}:
		Relative errors $\left|\Delta F/F\right|$ for $w_\phi(z)$, $\Omega_\mathrm{m}(z) = 1 - \Omega_\phi(z)$,
$E_\phi(z)$. 
The dotted curves depict the relative errors for the
$w_\phi \approx [0/1]_{w_\mathrm{DE}}(T(z))$ based approximations following from~\eqref{trackapprox},
while the thick dotted curves represent the relative errors for the $w_\phi \approx [1/1]_{w_\mathrm{DE}}(T(z))$
based approximations in~\eqref{trackapprox}. 
The dashed-dotted curves correspond to the CPL based approximations with, at the present time, numerically calculated
values $(w_0,w_a)=(-0.7581,0.1080)$. 
In Figure~(a) the space-dashed and dashed curves describe the Chiba approximation~\cite{chi10}, given in~\eqref{TwvarphiChiba},
truncated at 1st and 2nd order in $\Omega_\phi$, respectively. In Fig. (b) and (c) the dashed curves depict the
$w$CDM solution given in~\eqref{OmwEw}, which form the basis for the Chiba approximations.
}
	\label{Fig_RPPotApprox12}
\end{figure}

For numerical calculations we use the system~\eqref{uvsys} and the value
$\Omega_{\phi 0}=0.68$, which form a surface in the $(\tilde{\phi},u,v)$
state space that intersects the tracker orbit at a point found
numerically, thus identifying where $N=0$, and where the numerical calculations also give us
$w_\phi = u^2-1$, $\Omega_\mathrm{m} = 1 - \Omega_\phi = 1 - 3\tilde{\phi}v^2$,
while $E_\phi$ is found numerically by integrating\footnote{The present model, like all scalar field models,
also admits a 1-parameter set of thawing quintessence attractor solutions in the state space,
see~\cite{alhetal24}, which we for brevity do not discuss.}
\begin{equation}
E_\phi^\prime = - \frac32\left(1 + w_\mathrm{tot}\right)E_\phi
= - \frac32\left(1 + w_\phi\Omega_\phi\right)E_\phi
= - \frac32\left[1 + 3\tilde{\phi}v^2(u^2 - 1)\right]E_\phi.
\end{equation}
Figure~\ref{Fig_RPPotApprox12} depicts the relative errors of our approximations and
the Chiba approximations~\cite{chi10} for $\Omega_\mathrm{m}$ and $E_\phi$. Notably the
Chiba approximations are rather poor, even when compared to the $w_\phi \approx [0/1]_{w_\mathrm{DE}}(T(z))$
based approximation, which also have relative errors of several $\%$, while the $w_\phi \approx [1/1]_{w_\mathrm{DE}}(T(z))$
based approximations again are excellent with, e.g., a relative error of $\sim 0.1\%$ for $E_\phi$.
This is also the case for the CPL approximation, but in contrast to our
purely analytic approximations the parameters $w_0$ and $w_a$
were fitted numerically at the present time.
If we instead of using $\Omega_{\phi 0}$ in $T_0$
use the numerically computed $w_0 = w_{\phi 0}$ in $T_0$ as described in
eq.~\eqref{w11results} we get even more accurate results than our previous
approximation and the CPL approximation, but then this results in that the analytic
predictions of our approximations are lost.

Finally, note that decreasing $p$ to obtain observationally viable models yields even better results
since the maximal errors for the $w_\phi \approx [1/1]_{w_\mathrm{DE}}(T(z))$ based approximations are roughly
proportional to $p$, e.g. $p=0.1$ yields a maximal relative error $\sim 0.01\%$ for $E_\phi$.
Thus, again the $w_\phi \approx [1/1]_{w_\mathrm{DE}}(T)$ based approximations are observationally
indistinguishable from numerical results.

We then note that for $p=1$, our numerical calculations yield $(w_0,w_a)\approx(-0.7581,0.1080)$, and
hence $w^{\mathrm{CPL}}_\infty \approx-0.6501 $ and $\frac{w_a}{w_0 - w_\infty} \approx -1.1812$,
where we recall $w_\infty=-2/3\approx-0.6667$. The $w_\phi \approx [1,1]_{w_\mathrm{DE}}$
approximation yields $(w_0,w_a)\approx(-0.7519,0.0978)$, and hence
$\frac{w_a}{w_0 -w_\infty} \approx -1.1474$. Decreasing $p$ yields even better results.

\begin{figure}[ht!]
	\begin{center}
		\subfigure[Absolute difference of relative errors of $w_{\phi}$ for the CPL parametrisation]{\label{RPPot_Difwde_RelErrors}
			\includegraphics[width=0.31\textwidth]{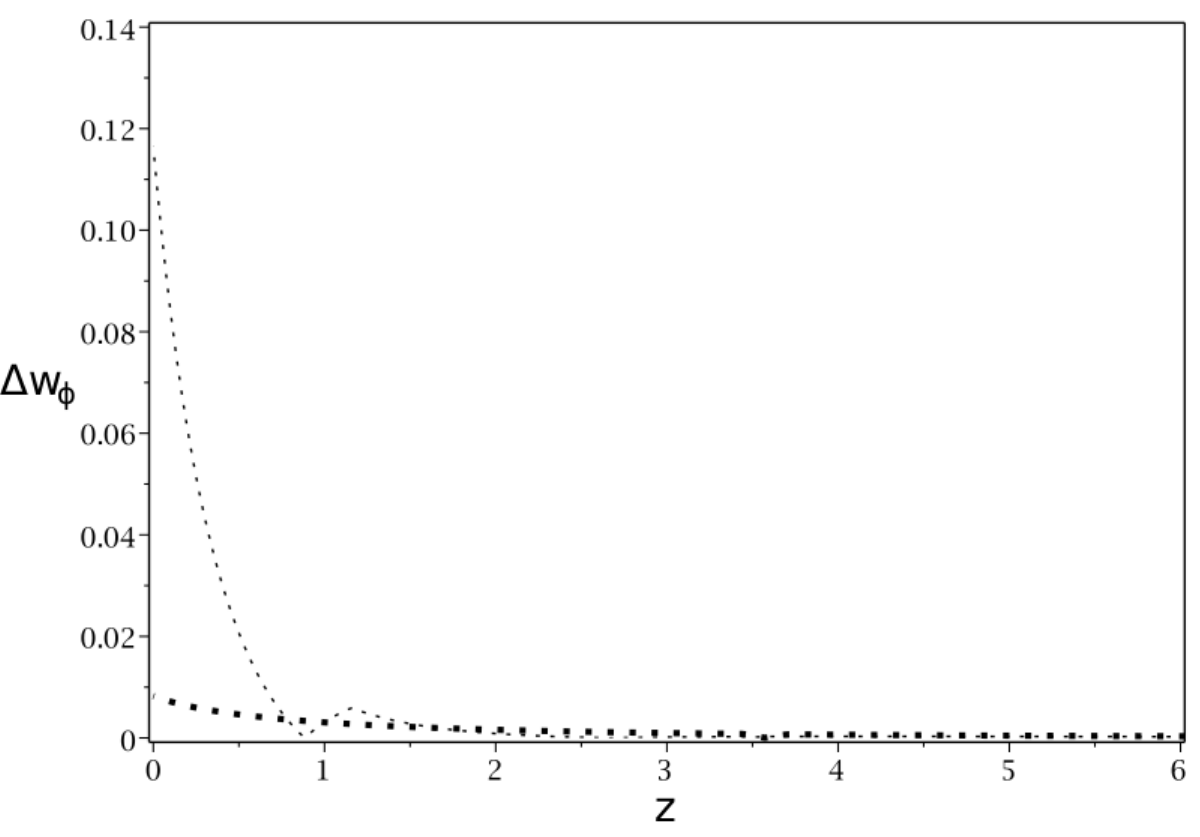}}
		\subfigure[Absolute difference of relative errors of $\Omega_\mathrm{m}$ for the CPL parametrisation]{\label{RPPot_DifOmm_RelErrors}
			\includegraphics[width=0.31\textwidth]{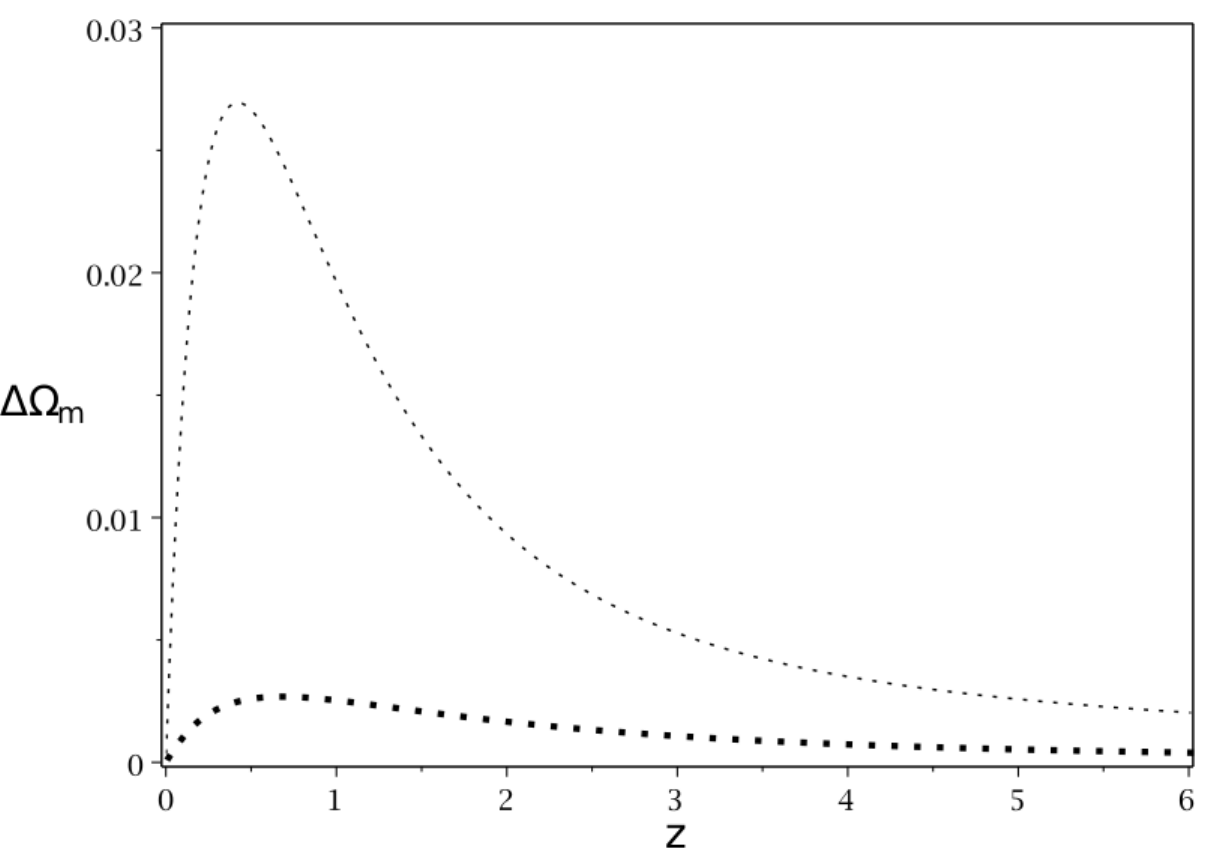}}
		\subfigure[Absolute difference of relative errors of $E_\phi$ for the CPL parametrisation]{\label{RPPot_DifE_RelErrors}
			\includegraphics[width=0.31\textwidth]{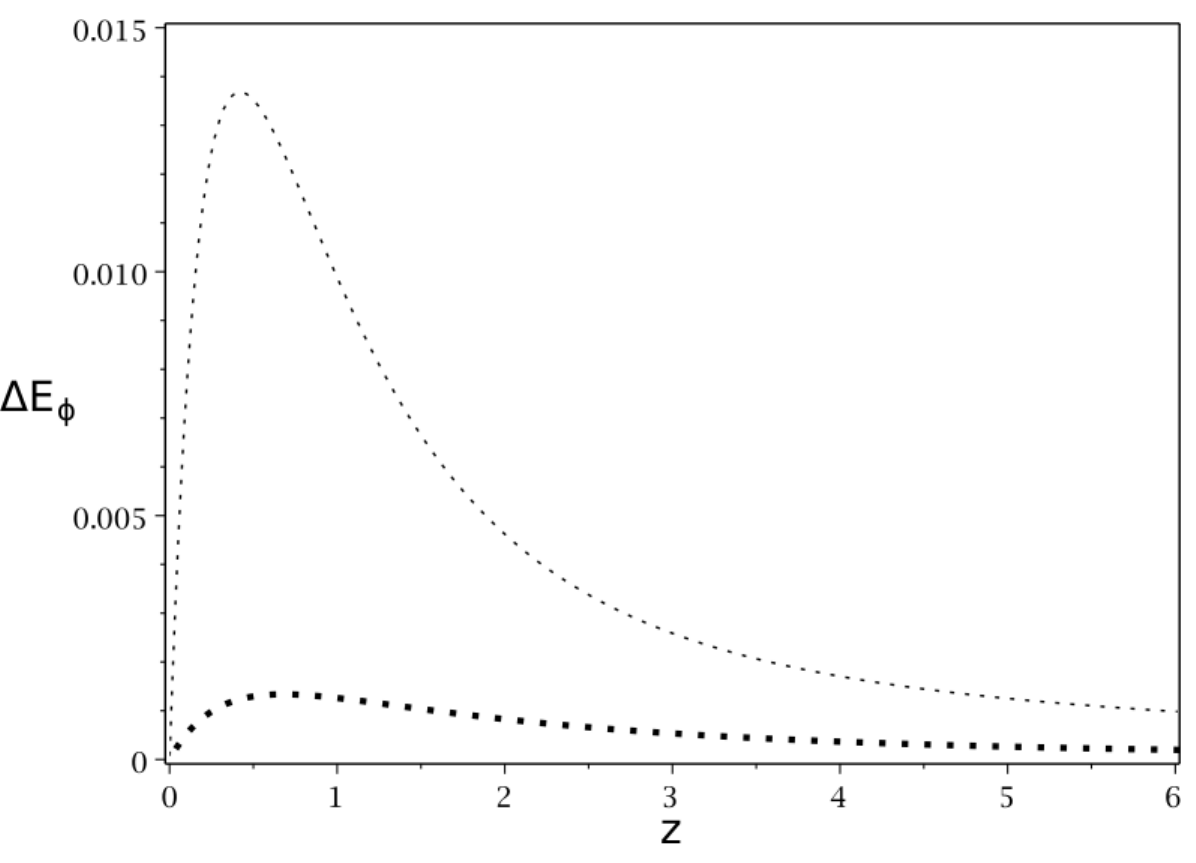}}
		\vspace{-0.5cm}
	\end{center}
	\caption{\emph{Tracking quintessence}: Absolute difference between relative errors, as
described in equations~\eqref{CPLcomperrors}, for the
CPL parametrisation relative to the numerical solution and the $w_\phi \approx [0/1]_{w_\mathrm{DE}}(T(z))$
(dotted curves) and $w_\phi \approx [1/1]_{w_\mathrm{DE}}$ (thick dotted curves) based approximations
for $w_\phi(z)$, $\Omega_\mathrm{m}(z)$, $E_\phi(z)$,
emphasizing that the analytic $w_\phi \approx [1/1]_{w_\mathrm{DE}}(T(z))$ based approximations are
observationally indistinguishable from numerical calculations, making the latter unnecessary.
	}
	\label{Fig_RPPotApprox}
\end{figure}

Figure~\ref{Fig_RPPotApprox} contains the differences $\Delta(w_\phi(z))$,
$\Delta(\Omega_\mathrm{m}(z))$ and $\Delta(E_\phi(z))$ between relative CPL errors
and the relative difference between CPL and our approximations, as
described in eq.~\eqref{CPLcomperrors}.
Again, the simple $w_\phi \approx [0/1]_{w_\mathrm{DE}}(T(z))$ approximant based approximations yields
order of estimate descriptions of the CPL approximations but not good enough for a detailed description;
on the other hand, our analytic $w_\phi \approx [1/1]_{w_\mathrm{DE}}(T(z))$ approximant based
approximations are very close to the numerical results, again illustrating that our best approximations
are so good so that they make numerical calculations irrelevant.

\section{Discussion\label{sec:disc}}

In this paper we have derived simple and accurate approximations for
thawing and tracking quintessence.
This was accomplished by identifying that thawing and tracking quintessence
are associated with `attractor solutions' for which
$\lim_{N\rightarrow -\infty}\Omega_\phi = 0$ (and hence
$\lim_{N\rightarrow -\infty}\Omega_\mathrm{m} = 1$) and
$\lim_{N\rightarrow -\infty}w_\phi = w_\infty$, where $w_\infty = -1$
for thawing quintessence while $-1< w_\infty <0$ for tracking quintessence.
These asymptotic properties served as motivation for situating the
derivation of the quintessence approximations in a unifying DE context.
It was shown that the conditions $\lim_{N\rightarrow -\infty}\Omega_\phi = 0$
and $\lim_{N\rightarrow -\infty}w_\phi = w_\infty \in [-1,0)$ naturally
resulted in series expansions in $T = T_0\exp(-3w_\infty N)$. However,
since our goal was to obtain accurate approximations during the whole time
period between matter dominance, i.e. when $\Omega_\phi\approx 0$ and
$\Omega_\mathrm{m}\approx 1$, and the present time for thawing and tracking
quintessence we used Pad\'e approximants where the $[1/1]_{\Omega_\mathrm{DE}}$ Pad\'e approximant
for $\Omega_\mathrm{DE}$ was used to relate the approximations to the present
time initial value $\Omega_{\mathrm{DE}0}$, or, equivalently,
$\Omega_{\mathrm{m}0} = \rho_{\mathrm{m}0}/3H_0^2$.

After the unifying DE section we specialized to thawing and tracking quintessence
by deriving two dynamical systems adapted to the two different types of quintessence.
This allowed us to insert the DE series expansions into these two dynamical systems
and solve for $\gamma$ and $\beta$, which then via the
$[0/1]_{w_\mathrm{DE}}(T(z))$ and $[1/1]_{w_\mathrm{DE}}(T(z))$ based DE expressions,
and changing $\Omega_{\mathrm{DE} 0}$ to $\Omega_{\phi 0}$, yield corresponding
quintessence approximations for $\Omega_\phi(z)$ and $E(z)=H(z)/H_0$.
Note that for a given potential, the parameters in our approximations precisely
correspond to the dimensionless parameters in the potential, since they are
easily analytically computed from the potential by taking scalar field derivatives;
thus, no extra parameters occur in our approximate expressions, although note that
thawing quintessence, which exists for \emph{all} scalar field potentials,
also involves a 1-parameter set of solutions described
by the freezing value $\phi_*$ during the matter dominated epoch. We also showed
numerically that the $w_\phi(z) \approx [1/1]_{w_\mathrm{DE}}(T(z))$ based approximations
are an order of magnitude more accurate than any previous analytic approximations
in the literature, making numerical calculations for thawing and tracking quintessence
unnecessary. Moreover, we showed how one can \emph{analytically} compute the CPL parameters,
again with an accuracy that eliminates the need for numerical calculations of these
parameters.

In this paper we have followed the majority of work in this area by focussing on
generating results from $w_\phi$, but there are many other possibilities,
as well as further developments and applications, where we describe
some in the following list:
\begin{itemize}
\item Instead of using $w_\phi$ as the starting point, one can use
other variables. For example, as pointed out in footnote~\ref{for:E2}, our
series expansions also give rise to a series expansion in $T$ for
$E_\mathrm{DE}^2$ and thereby also for $E_\mathrm{DE}$ and $E_\phi$. This
expansion can subsequently be used to obtain various Pad\'e approximants,
which can then serve as the starting point for calculating other
quantities. Or one can derive series expansions for one of the distance measures
and use this for constructing Pad\'e approximants, from which
other variables and observables can be derived.
\item We have here neglected radiation, which, however, can be included in
several different ways; one can e.g. use the fixed points on the radiation
boundary as the starting point instead of the fixed points on the matter
boundary when it comes to quintessence, as was done
in~\cite{alhugg23}. Or one can simply add radiation,
$\rho_\gamma = \rho_{\gamma 0}\exp(-4N) = 3H_0^2\Omega_{\gamma 0}\exp(-4N)$, to
$\rho$, i.e. $\rho = \rho_\gamma + \rho_\mathrm{m} + \rho_\mathrm{DE}$, and
keep the present $w_\mathrm{DE}$ expressions, since radiation has a negligible
influence on quintessence evolution, and, e.g., use that $3H^2 = \rho$.
\item For some purposes there might be of interest to obtain an expression for how the
scalar field changes during thawing quintessence from the past freezing value $\phi_*$
during the matter dominant epoch
to some arbitrary $e$-fold time $N$. This can be done in several ways, but constructing
the $[1,1]_{\phi^\prime}$ Pad\'e approximant on the right hand side of $\phi^\prime$
in~\eqref{phipT}, i.e.
$\sqrt{3(\gamma-1)}\, T/(1 + \sigma T)$ where
$\sigma = \frac{4}{5}\left(1+\frac{\eta_*}{6}\right)$,
and then integrating from $-\infty$ to $N$ yields an excellent
approximation\footnote{It is, e.g., a better approximation than that of
Raveri {\it et al.}~\cite{ravetal19} given in
equation~\eqref{Raveri} in appendix~\ref{app:quintlit}.}
given by
\begin{equation}\label{FLVarphiInt}
\phi = \phi_* + \sqrt{\sfrac13(\gamma-1)}\,\left(\frac{1}{\sigma}\right)
\ln{\left(1+\frac{\sigma\Omega_{\phi 0}}{(1 - \gamma\Omega_{\phi 0})e^{-3N}}\right)}.
\end{equation}
\item One can use other dynamical systems variables
than the present ones to obtain somewhat different results.
For example, we could have used the variables in~\cite{alhetal24} to deal with tracking
quintessence, which would have been useful for the hyperbolic potentials that in addition
to the inverse power law potential was dealt with in that paper.\footnote{This would
had come with the price that $\Gamma$ would not have appeared explicitly in the results
for $\gamma$ and $\beta$ due to the different regularization of $\lambda$.}
However, to improve convergence one should then replace the scalar field variable
$\bar{\varphi}$ in~\cite{alhetal24} with $\tilde{\varphi} = \bar{\varphi}^2$.
\item In the spirit of CPL, it is possible to improve our quintessence approximations
even further by viewing $\gamma$ and $\beta$ as numerically determined parameters.
This, however, means that one (similarly to CPL) would only do curve fitting,
which implies less theoretical content and predictive power.
\item We have here focussed on obtaining thawing and tracking quintessence approximations
situated in a broader unifying DE context for spatially homogeneous, isotropic and flat spacetimes.
A natural next step is to use these approximations in order to interpret and assess
the rapidly growing set of increasingly precise observational data. In particular, the
present results, or variations thereof, can serve as input in cosmological perturbation
theory to bring the results closer to a wide set of observations.
\item We finally note that the present approach and variations thereof can easily be
applied to other models and theories, especially for those admitting an Einstein frame,
although this usually adds parameters that diminish predictive power.
\end{itemize}
\subsection*{Acknowledgments}
A. A. is supported by FCT/Portugal through CAMGSD, IST-ID, projects UIDB/04459/2020
and UIDP/04459/2020, and by the H2020-MSCA-2022-SE project EinsteinWaves, GA No. 101131233.
A.A. would also like to thank the CMA-UBI in Covilh\~a for kind hospitality.
C. U. would like to thank the CAMGSD, Instituto Superior T\'ecnico in Lisbon for kind hospitality.
%

\begin{appendix}

\section{Quintessence approximations and comparisons\label{app:quintlit}}

In this appendix we review the most prominent quintessence approximations in
the literature, including some simplified derivations and extensions;
moreover, we will present them in a more unified and concise manner
than their original form. We begin with thawing quintessence.

\subsection{Thawing quintessence}

Let us first derive an approximation introduced by Sherrer and Sen (2008)
in~\cite{schsen08} and further explored by e.g. Agrawal {\it et al.}~\cite{agretal18}
and Raveri {\it et al.}~\cite{ravetal19}. Instead of using
the state vector $(\phi,\Sigma_\phi,\Omega_\phi)$
we construct a dynamical system closely connected with $\tilde{\phi},u,v$ based on
the state vector $(\phi,u,\omega)$ where\footnote{The variable $u$ was introduced in~\cite{alhetal23}
while the variable $v$ in that paper is just $v = \omega/\sqrt{3}$.}
\begin{equation}
u = \Sigma_\phi\sqrt{\frac{2}{\Omega_\phi}} = \frac{\phi^\prime}{\sqrt{3\Omega_\phi}},\qquad
\omega = \sqrt{\Omega_\phi},
\end{equation}
which due to~\eqref{Dynsysthaw} yields the dynamical system
\begin{subequations}\label{Dynsyssom}
\begin{align}
{\phi}^\prime &= \sqrt{3}u\omega,\label{varphiprimeuomega}\\
u^\prime &= \frac12(2 - u^2)[\sqrt{3}\lambda({\phi})\omega - 3u],\\
\omega^\prime &= \frac32(1-u^2)\omega(1-\omega^2).
\end{align}
\end{subequations}

Next $\phi$ is assumed to take a frozen constant value $\phi = \phi_*$
while the two last equations in~\eqref{Dynsyssom} are linearized in
$u$ \emph{only}, which results in the following equations for $u$ and $\omega$:
\begin{subequations}
\begin{align}
u^\prime &\approx - 3u + \sqrt{3}\lambda_*\omega,\label{uapprox2}\\
\omega^\prime &\approx \frac32\omega(1-\omega^2),\label{omegaeq}
\end{align}
\end{subequations}
where $\lambda_* = \lambda({\phi}_*)$. The second equation,
which corresponds to
$\Omega_\phi^\prime \approx 3\Omega_\phi(1-\Omega_\phi)$ and hence
$[\ln\{\Omega_\phi/(1-\Omega_\phi)\}]^\prime = [\ln(\Omega_\phi/\Omega_\mathrm{m})]^\prime
= [\ln(\rho_\phi/\rho_\mathrm{m})]^\prime\approx 3$,
where $\rho_\mathrm{m} = \rho_{\mathrm{m},0}\exp(-3N)$ yields $\rho_\phi \propto \mathrm{constant}$
and thereby the $\Lambda$CDM expression for $\omega(N)$. 
%
%
Inserting this solution into~\eqref{uapprox2} and integrating gives
$u(N)$, from which $w_\phi = u^2 - 1$ follows;
deparameterizing the solution results in $u(\omega)$ and thereby $w_\phi(\omega)$;
alternatively, solve directly
$du/d\omega = u^\prime/\omega^\prime \approx
-\frac{2u}{\omega(1-\omega^2)} + \frac{2\lambda_*}{\sqrt{3}(1-\omega^2)}$:
\begin{subequations}\label{swappr}
\begin{align}
\begin{split}
1 + w_\phi &= \frac{\lambda_*^2}{3}\left[\omega^{-1} - (\omega^{-2} - 1)\tanh^{-1}(\omega)\right]^2\\
&=
\frac{\lambda_*^2}{3}\left[\Omega_\phi^{-1/2} - (\Omega_\phi^{-1} - 1)\tanh^{-1}(\Omega_\phi^{1/2})\right]^2 \label{wvarphiSherrer},
\end{split}\\
\Omega_\phi &= \left[1 + \left(\Omega_{\phi 0}^{-1} - 1\right)e^{-3N}\right]^{-1},\label{OmvarphiN}
\end{align}
\end{subequations}
where the last expression coincides with that of $\Lambda$CDM when identifying
$\Omega_{\phi}$ with $\Omega_{\Lambda}$ (this approximation thereby
does not give thawing corrections to the line element since it is just the
$\Lambda$CDM line element); the above expressions
summarize eqs. (23), (25) and (26) in Scherrer and Sen~\cite{schsen08}.
Taken together with eq.~\eqref{varphiprimeuomega} and $u^2 = 1 + w_\phi$ this results in
\begin{equation}\label{Sherrer2}
\phi^\prime = \lambda_*\left[1 - (\omega^{-1}- \omega)\tanh^{-1}(\omega)\right] =
\lambda_*\left[1 - (\Omega_\phi^{-1/2}- \Omega_\phi^{1/2})\tanh^{-1}(\Omega_\phi^{1/2})\right],
\end{equation}
where inserting~\eqref{OmvarphiN} yields $\phi^\prime(N)$. This expression agrees with
eq. (15) in Agrawal {\it et al.}~\cite{agretal18} for the exponential potential for which $\lambda$
is a constant. 
Integration, with the condition that ${\phi} = {\phi}_*$
when $N\rightarrow - \infty$ and $\Omega_\varphi \rightarrow 0$ (using
$\phi^\prime = \Omega_\phi^\prime d\phi/d\Omega_\phi$ and
$\Omega_\phi^\prime \approx 3\Omega_\phi(1 - \Omega_\phi)$),
results in
\begin{equation}\label{Raveri}
\phi \approx 
\phi_* + \frac{2\lambda_*}{3}\left[\Omega_\phi^{-1/2}\tanh^{-1}(\Omega_\phi^{1/2}) - 1\right],
\end{equation}
where $\phi(N)$ is obtained by inserting~\eqref{OmvarphiN}. This approximation
coincides with eq. (8) in Raveri {\it et al.}~\cite{ravetal19}.

%

Other types of thawing quintessence approximations in the literature are based
on series expansions of $V(\phi)$ at some $\phi_*$, and thus also of
$\lambda(\phi)$ when the latter is regular at $\phi_*$. An example of this were given by
Cahn {\it et al.} (2008)~\cite{cahetal08} who gave approximate expressions for
$\phi^\prime$ (replacing $\dot{\phi}$ with $\phi^\prime$ in their eq. (15))
and $\phi$, although note that $\phi_*$ is missing in their
eq. (16), thus hiding that you typically are interested in a domain of $\phi_*$.
These expressions correspond to the present second order expressions in
eqs.~\eqref{phiT} and~\eqref{phipT}, whose accuracy we subsequently improved
by taking Pad\'e approximants.

Based on earlier results by Dutta and Scherrer (2008)~\cite{dutsch08}, Chiba (2009)~\cite{chi09}
Taylor expanded the potential to second order in $\phi-\phi_*$.
By a sequence of \emph{ad hoc} approximations in the proper time $t$
the author derived a series of results (see p. 083517-4 in~\cite{chi09}),
which can be extended to also include $\phi^\prime$ and $w_\mathrm{tot}$, more simply
expressed in $\omega=\sqrt{\Omega_\phi}$ as follows:
\begin{subequations}\label{Kappreq}
\begin{align}
\phi  &=\phi_*+\frac{2\lambda_*}{3}\left(\frac{(1-\omega^2)^{\frac{1}{2}(1-K)}\left[(1+\omega)^K-(1-\omega)^K\right]-2K\omega}{K\omega(K^2-1)}\right), \label{Chiba1} \\
\phi^\prime &=\lambda_* \frac{(1-\omega^2)^{\frac{1}{2}(1-K)}\left[(1+K\omega)(1-\omega)^K-(1-K\omega)(1+\omega)^K\right]}{K\omega(K^2-1)}, \label{Chiba2}\\
1+ w_\phi &=\frac{\lambda^2_*}{3}(1-\omega^2)^{1-K}\left[\frac{(1+K\omega)(1-\omega)^K-(1-K\omega)(1+\omega)^K}{K\omega^2(K^2-1)}\right]^2, \label{wvarphiChiba}\\ 
w_\mathrm{tot} &= w_\phi\omega^2,
\end{align}
\end{subequations}
where
\begin{equation}
K = \sqrt{1 - \frac43 \eta_*,
}
\end{equation}
while $\omega = \sqrt{\Omega_\phi}$ is given by its $\Lambda$CDM expression
\begin{equation}\label{OLambda}
\omega = \left(1 + (\Omega_{\phi 0}^{-1} - 1)e^{-3N}\right)^{-1/2}.
\end{equation}
The right hand side for $\phi^\prime$ was obtained by using
$\phi^\prime = \frac{d\phi}{d\omega}\omega^\prime$
and the $\Lambda$CDM relation $\omega^\prime = \frac32(1 - \omega^2)\omega$.
Taking the limit $K\rightarrow 1$ of equations~\eqref{Chiba1} and \eqref{Chiba2},
which corresponds to neglecting the second order Taylor
expansion term of the potential, i.e. setting $V_{,\phi\phi}(\phi_*)=0$,
leads to an indeterminacy that can be solved by L'H\^opital's rule.
In~\cite{chi09} the author also showed how to convert the expression for $w_\phi$
to the case when $K^2<0$, which resulted in the approximation
\begin{equation}\label{K2neg}
1+w_\phi=\frac{4}{3}\lambda^2_* (1-\omega^2)\left[\frac{\tilde{K}\omega\cos{\left(\tilde{K}\sinh^{-1}{\left(\frac{\omega}{\sqrt{1-\omega^2}}\right)}\right)}
- \sin{\left(\tilde{K}\sinh^{-1}{\left(\frac{\omega}{\sqrt{1-\omega^2}}\right)}\right)}}{\tilde{K}\omega^2(1+\tilde{K}^2)}\right]^2,
\end{equation}
where
\begin{equation}
\tilde{K} = \sqrt{\frac43 \eta_*-1
}.
\end{equation}
In addition, Chiba~\cite{chi09} discussed
the approximation given by Crittenden {\it et al.} (2007)~\cite{crietal07}.
\subsection{Tracking quintessence}

Here we consider the tracking quintessence approximations given by Chiba (2010)~\cite{chi10},
which in turn relied on earlier work by Watson and Scherrer (2003)~\cite{watsch03},
for the case $\Gamma = \mathrm{constant}$, and, in particular,
$\Gamma = \mathrm{constant}>1$, which yields the inverse power law potential.
This work is based on a linearized equation for a perturbation of $w_\phi$ with
respect to its asymptotic value $w_\infty$. To obtain an improved
approximation for the \emph{linearized solution} for $w_\phi$ Chiba~\cite{chi10}
considered a series expansion in $e^{-3w  N}$, when expressed in
$e$-fold time and the present notation. This is the same series expansion quantity as
in the present paper, but in contrast to Chiba~\cite{chi10} we used it to produce approximations
for the full \emph{non-linear} equations; thus, the expansions in the present work coincide with those
in~\cite{watsch03} and~\cite{chi10} for the quantities that have been computed in those
two references up to first order
but not for higher orders since Chiba~\cite{chi10} only deals with the linearized equation for $w_\phi$.
Moreover, there is translation freedom in $N$. In the present work this freedom
was fixed by the $[1/1]_{\Omega_\phi}$ Pad\'e approximant for $\Omega_\phi$,
which resulted in an expansion in
$T = \Omega_{\phi 0}e^{-3w_\infty N}/(1 - \gamma\Omega_{\phi 0})$; Chiba~\cite{chi10}
on the other hand fixed this freedom by, in effect, setting $w_\phi = w_\infty$,
to obtain the following approximation for $\Omega_\phi$:
\begin{equation}\label{OmstarChiba}
\Omega_\phi \approx \frac{\tilde{T}}{1+\tilde{T}} =
\frac{\Omega_{\phi 0}}{\Omega_{\phi 0} + (1 - \Omega_{\phi 0})e^{3w_\infty N}}, 
\qquad
\tilde{T} = \left(\frac{\Omega_{\phi 0}}{1 - \Omega_{\phi 0}}\right)e^{-3w_\infty N},
\end{equation}
which corresponds to $w$CDM, while the series expansion
in $\tilde{T}$ for the linearized equation for $w_\phi$ resulted in
\begin{equation*}
w_\phi = w_\infty + w_\infty(1-w_\infty^2)\sum_{n=1}^\infty
\left(\frac{(-1)^{n-1}}{2n(n+1)w_\infty^2 - (n+1)w_\infty + 1}\right)\tilde{T}^n.
\end{equation*}
Note that the quantity $\tilde{T}$ is related to $T$ according to
\begin{equation*}
\tilde{T} = \left(\frac{1 - \gamma\Omega_{\phi 0}}{1 - \Omega_{\phi 0}}\right)T,
\end{equation*}
which corresponds to a translation in $N$ with the constant
$-(1/3w_\infty)\ln[(1 - \gamma\Omega_{\phi 0})/(1 - \Omega_{\phi 0})]$.

As a next step Chiba~\cite{chi10} replaced $\tilde{T}$ with
$\tilde{T} = \Omega_\phi/(1 - \Omega_\phi)$,
%
%
and expanded the resulting expression for $w_\phi$ in $\Omega_\phi$, which
yields the following series, here truncated at second order, in $\Omega_\phi$:
\begin{equation}\label{TwvarphiChiba}
w_\phi \approx w_\infty +
\left(\frac{w_\infty(1-w_\infty^2)}{4w_\infty^2 - 2w_\infty + 1}\right)\Omega_\phi
\left[1 +
\left(\frac{w_\infty(8w_\infty-1)}{12w_\infty^2 - 3w_\infty + 1}\right)\Omega_\phi\right].
\end{equation}
Note that this is an expansion that is designed to improve the solution for the
\emph{linearized} equation for $w_\phi$, \emph{not} the actual solution of the
\emph{non-linear} equation for $w_\phi$. The improved
linear approximation only improves the actual solution to the extent the linear approximation describes that solution,
which, fortunately, to some extent is the case for the inverse power law potential,
see e.g., Figure 1 in~\cite{tsu13}. However, not surprisingly, these approximations yield
poorer accuracy than the present ones that are based on approximating the full non-linear equations.
\section{Scaling freezing quintessence\label{app:scaling}}

In~\cite{alhetal23} it was demonstrated that in appropriate variables, resulting
in a global 3D state space setting, scaling freezing quintessence is associated with a
single attractor orbit that constitutes the unstable manifold of an isolated
fixed point. The situation and key attractor mechanism is thereby the same as for
tracking quintessence. An open set of nearby orbits to the (hyperbolic)
saddle fixed point is pushed toward the attractor orbit by a 2D invariant stable
manifold boundary set, for details, see~\cite{alhetal23}
and~\cite{alhetal24}. The difference with tracking quintessence is that
scaling freezing quintessence arises from potentials that have an
asymptotic exponential behaviour that is sufficiently steep.

We therefore consider
the situation where $\lim_{\phi\rightarrow -\infty}\lambda = \lambda_-$,
$\lambda_- > \sqrt{3}$, but where compatibility with nuclear synthesis data
requires $\lambda_- \gg \sqrt{3}$, see~\cite{alhetal23} and references therein.
Let us further assume that the scalar field, during the time epoch
$N\in (-\infty,0]$ for the attractor solution, rolls down a scalar field
potential with $\lambda(\phi)>0$ where $\lambda(\phi)$ is monotonically decreasing, i.e.,
we assume that
\begin{equation}\label{GammaDef}
\Gamma = \frac{V\,V_{,\phi\phi}}{V_{,\phi}^2} = 1 + (\lambda^{-1})_{,\phi} > 1 \quad \text{and} \quad
\lim_{\phi\rightarrow -\infty}\Gamma = 1.
\end{equation}

For our analysis we will consider the dynamical system for the state space
vector $(\lambda, \Sigma_\phi, \Omega_\phi)$, which can be
obtained from~\eqref{Dynsysthaw}:
\begin{subequations}\label{DynsysScal}
\begin{align}
\lambda^\prime &= -\sqrt{6}(\Gamma - 1)\lambda^2 \Sigma_\phi, \\
\Sigma_\phi^\prime &= - \frac32(1 + \Omega_\phi - 2\Sigma_\phi^2)\Sigma_\phi +
\sqrt{\frac32}\,\lambda\,(\Omega_\phi - \Sigma_\phi^2), \label{Sigmaeqmatter3} \\
\Omega^{\prime}_\phi &= 3(\Omega_\phi - 2\Sigma_\phi^2)(1-\Omega_\phi),\label{Omphieq3}
\end{align}
\end{subequations}
where $\Gamma$ now is considered to be a function of $\lambda$, i.e., $\Gamma = \Gamma(\lambda)$.

Before continuing, let us note that the simple example of a potential consisting of two
exponential terms,
\begin{equation}
V = V_-e^{-\lambda_-\phi} + V_+e^{-\lambda_+\phi}, \qquad \lambda_->0,\qquad \lambda_- > \lambda_+,\qquad V_\pm >0,
\end{equation}
yields
\begin{equation}\label{2expGamma}
\Gamma = 1 + \frac{(\lambda_- - \lambda)(\lambda - \lambda_+)}{\lambda^2}.
\end{equation}

In the present variables, scaling freezing quintessence is associated with
the unstable manifold orbit that originates from the fixed point
\begin{equation}
\mathrm{S}\!:\quad (\lambda,\Sigma_\phi,\Omega_\phi) =
\left(\lambda_-,\sqrt{\frac{3}{2}}\left(\frac{1}{\lambda_-}\right),\frac{3}{\lambda^2_-}\right).
\end{equation}
Next we assume that $\Gamma(\lambda)$ can be Taylor expanded around $\lambda_-$, i.e.,
\begin{equation}
\Gamma = 1 + \Gamma_{1}(\lambda - \lambda_-)+\frac{\Gamma_2}{2}(\lambda - \lambda_-)^2 + \dots,\qquad
\Gamma_n = \left.\frac{d^{n}\Gamma}{d\lambda^n}\right|_{\lambda=\lambda_-}.
\end{equation}
Note that~\eqref{2expGamma} is regular since $\lambda > \lambda_- >0$ for the potential
with two exponential terms and that
\begin{equation}
\Gamma_1 = -\left(\frac{\lambda_- -\lambda_+}{\lambda_-^2}\right),\qquad
\Gamma_2 = \frac{2(\lambda_- - 2\lambda_+)}{\lambda_-^3}
\end{equation}
in this case.

We can now proceed in the same manner as for thawing and tracking quintessence and
first make an expansion based on the unstable eigenvalue, $-3\lambda_-\Gamma_1$, which
leads to expansions in
\begin{equation}
T = C e^{-3\lambda_-\Gamma_1 N}.
\end{equation}
The fact that $\lim_{N\rightarrow - \infty}\Omega_\phi >0$ and that the eigenvalue
involves two parameters, $\lambda_-$ and $\Gamma_1$, result in quite messy expressions
for the series expansions. For brevity we will therefore truncate the series already
at the linear order in $T$:
\begin{subequations}
\begin{align}
\lambda &= \lambda_- - T + \dots, \\
\Sigma_\phi &= \sqrt{\frac{3}{2}}\left(\frac{1}{\lambda_-}\right)\left(1 + \alpha T + \dots\right), \\
\Omega_\phi &= \frac{3}{\lambda^2_-}\left(1 + \beta T + \dots\right),
\end{align}
\end{subequations}
where
\begin{equation}
\alpha=\frac{\lambda^3_-\Gamma_1+\lambda^2_- -3}{\lambda_-(2\lambda^4_- \Gamma^2_1-\lambda^3_-\Gamma_1+\lambda^2_--3)}, \qquad
\beta = \frac{2(\lambda^2_--3)}{\lambda_-(2\lambda^4_- \Gamma^2_1-\lambda^3_-\Gamma_1+\lambda^2_--3)}.
\end{equation}
Using that $w_\phi = -1 + 2\Sigma_\phi^2/\Omega_\phi$ yields
\begin{equation}
w_\phi = (2\alpha-\beta)T + \dots,
\end{equation}
where
\begin{equation}
2\alpha-\beta = \frac{2\lambda^2_-\Gamma_1}{\lambda_-(2\lambda^4_- \Gamma^2_1-\lambda^3_-\Gamma_1+\lambda^2_--3)}.
\end{equation}
Note that $\lim_{N\rightarrow - \infty} w_\phi =0$, which leads to that asymptotically
toward the past $\rho_\phi \propto \rho_\mathrm{m} \propto a^{-3}$,
hence the name \emph{scaling} freezing,
while $w_\phi$ decreases toward the future.

We leave the higher order series expansions, the construction of Pad\'e approximants,
the identification of $C$ with present day data, and subsequent investigations, such as
error estimates for various potentials, for the interested reader.

\end{appendix}

\bibliographystyle{unsrt}
\bibliography{../Bibtex/cos_pert_papers}

\end{document}